\documentclass[reprint,
superscriptaddress,
groupedaddress,
unsortedaddress,
runinaddress,
frontmatterverbose, 
nofootinbib,
nobibnotes,
bibnotes,
amsmath,amssymb,
aps,
floatfix,
]{revtex4-2}
\usepackage[dvipsnames]{xcolor}
\usepackage{graphicx,
subfig,
lipsum
}
\usepackage{hyperref}
\usepackage{nameref}
\usepackage[justification=raggedright,singlelinecheck=false]{caption}
\usepackage[labelfont=bf, font=small]{caption}
\usepackage[font=footnotesize]{caption}

\usepackage{float}
\usepackage{hyperref}
\usepackage{amsthm}  
\usepackage{amsmath} 

\usepackage{caption}
\usepackage{booktabs} 
\usepackage{algorithm}
\usepackage{pgfplots}
\pgfplotsset{compat=1.18}
\usepackage{algpseudocode}  
\usepackage{amsmath,amssymb}
\usepackage{braket}
\usepackage{graphicx}
\usepackage{stackengine}
\usepackage{bm}
\usepackage[utf8]{inputenc}
\usepackage{lmodern}
\usepackage{mathtools} 
\usepackage{tikz}
\usetikzlibrary{shapes}  
\usepackage{graphicx}
\usepackage{tabularray}
\usepackage[mathlines]{lineno}
\usepackage{subcaption}
\usepackage{fancyhdr}
\usepackage{caption}
\usepackage{babel}
\usepackage{csquotes}
\usepackage{natbib}

\begin{document}

\preprint{APS/123-QED}
\title{Supervised Machine Learning for Predicting Open Quantum System Dynamics and Detecting Non-Markovian Memory Effects}

\author{Ali Abu-Nada}
\email{ali.abunada@sma.ac.ae}
\affiliation{Sharjah Maritime Academy, Sharjah, United Arab Emirates}
\author{Subhashish Banerjee}
\email{s.banerjee@iitj.ac.in}
\affiliation{Department of Physics, Indian Institute of Technology Jodhpur, 342030, India}

\begin{abstract}
We present a \emph{novel} and scalable supervised machine learning framework to predict open-quantum system dynamics and detect non-Markovian memory using only local ancilla measurements. A system qubit is coherently coupled to an ancilla via a symmetric XY Hamiltonian; the ancilla interacts with a noisy environment and is the only qubit we measure. A feedforward neural network, trained on short sliding windows of supplementary data from the past, forecasts the observable system $\langle Z_{(S)}(t)\rangle$ without state tomography or knowledge of the bath.

To quantify memory, we introduce a normalized revival-based metric that counts upward 'turn-backs' in \emph{predicted} $\langle Z_{(S)}(t)\rangle$ and reports the fraction of evaluated samples that exceeds a small threshold. This bounded score provides an interpretable, model-independent indicator of non-Markovianity.

We demonstrate the method on two representative noise channels, non-unital amplitude damping and unital dephasing from random telegraph noise (RTN). Under matched conditions, the model accurately reproduces the dynamics and flags memory effects, with RTN exhibiting a larger normalized revival score than amplitude damping. Overall, the approach is experimentally realistic and readily extensible, enabling real-time, interpretable non-Markovian diagnostics from accessible local measurements.
\end{abstract}

\maketitle
\section{Introduction}\label{sec:sec1}

The study of open quantum systems is central to understanding realistic quantum dynamics, where no system is perfectly isolated. In practical settings, quantum systems interact with external environments, leading to noise, decoherence, and information loss~\cite{breuer2002theory, rivas, nielsen2010quantum, lidar,banerjee2018open,PhysRevA.110.052209}. These effects are especially significant in the field of quantum information science, where applications like quantum computation~\cite{nielsen2010quantum}, quantum communication~\cite{Gisin2007,pingpongsrikanth}, and quantum sensing~\cite{Degen2017} rely on maintaining fragile quantum coherence. Because of this, modeling and managing the influence of the environment is crucial.

The behavior of an open quantum system can generally be classified as either Markovian or non-Markovian. In Markovian dynamics, the environment acts like a memoryless sink: information flows out of the system and does not return. This leads to a smooth and irreversible decay of system observables, such as the qubit coherence or polarization \cite{banerjee2018open, nielsen2010quantum, abunada}. In contrast, non-Markovian dynamics involve memory effects, where information that left the system can flow back from the environment at a later time~\cite{breuer, rivas2, banerjee2018open,sss,sbdephasing,sbhieracrchy,PhysRevA.99.042128}. This results in non-monotonic behavior, such as temporary revivals in observables like the expectation value of a Pauli operator~\cite{liu}. These revivals serve as signatures of information backflow and are widely recognized as indicators of non-Markovianity.

Traditional techniques to detect non-Markovianity often involve measuring quantities like the trace distance between quantum states~\cite{laine2010measure}, examining CP-divisibility~\cite{rivas2}, analyzing the behavior of entanglement between subsystems~\cite{rivas},   using witness operators that detect backflow of information~\cite{maniscalco}, deviation from temporal self-similarity~\cite{sss}. While powerful, these approaches usually require access to the full quantum state or dynamical map, which demands full state tomography or repeated measurements. This process becomes impractical in experiments with many qubits or limited measurement access.

To address these challenges, researchers have developed alternative methods that rely solely on simple, experimentally accessible observables. For instance, experimental quantum probing measurements \cite{sbhenri1} and snapshot verification of non-Markovianity with unknown system-probe coupling \cite{sbhenri2}, mutual information flow~\cite{fanchini2014non}, quantum Fisher information~\cite{luo2012quantum,sbunruhfisher}, and engineered interference measurements~\cite{Cialdi2017} have all been explored as indirect indicators of memory effects. Among these strategies, continuously monitoring the expectation value of a single Pauli operator, such as $\langle Z(t) \rangle$, over time has emerged as a simple yet powerful way to reveal memory effects in practice~\cite{liu, Campbell}. This method is especially attractive because it only requires local, single-qubit measurements, making it both experimentally feasible and scalable to larger systems.

This work introduces a simple framework with two main goals. The first goal is to build a predictive model using \textit{supervised machine learning} (ML) to estimate how an open quantum system evolves over time. This is accomplished using only a short sequence of recent measurements on an ancilla qubit that interacts directly with the environment while being coherently coupled to the system qubit, thereby encoding information about the system’s dynamics. Our ML model, implemented here as a feedforward neural network~\cite{Goodfellow2016}, does not require full access to the system’s quantum state or detailed knowledge of the environment. Instead, it learns to predict the observable value of the system (specifically, the expectation value of the Pauli-$Z$ operator) by processing a sliding window of recent ancillary measurements. This approach enables us to model the behavior of the system using only partially experimentally realistic data.

The second goal follows directly from the first: having predicted the system observables, we introduce a revival-based method to detect and quantify non-Markovian memory effects. We leverage the fact that non-Markovian behavior manifests as temporary reversals (\textit{revivals}) in the system’s dynamics. We formulate a practical metric that detects these revivals in the predicted observables and assigns a bounded score reflecting the strength of memory effects. The procedure is intuitive and experimentally viable, requiring only local measurements and no full state reconstruction.

Our model is based on a simple but powerful idea: the system qubit interacts only with an ancilla qubit, and not directly with the environment. Instead, the ancilla is the one that connects to the environment and picks up its effects. This setup is known as a collision-model-based architecture~\cite{Ciccarello2013}, where either the ancilla interacts with many environmental parts (like a chain of collisions), or the environment indirectly affects the system through the ancilla. As a result, the system can still show memory effects---manifested as non-Markovian dynamics---even though it never directly interacts with the environment. These memory effects can appear as revivals in the system’s behavior over time, such as in the expectation value of a Pauli observable.\\
A similar idea was proposed theoretically in ~\cite{Campbell}, where a system is monitored using only an ancilla qubit that is coupled to the environment. It was shown that it is possible to learn about the environment’s influence without observing the system directly. Although this approach was theoretical, it supports the foundation of our supervised machine learning method, which also relies on observing only the ancilla.\\
More recently, this kind of setup has been tested in real experiments. In ~\cite{Gaikwad}, it was demonstrated using superconducting qubits, just like those used in IBM's quantum processors. In their experiment, two qubits were entangled in a Bell state. One of them acted as the ancilla and was connected to a small, engineered environment made from another qubit and a resonator. By tracking how entanglement changed over time, they observed its collapse and later revival---clear evidence of memory effects. They could even control the level of non-Markovianity by adding dephasing to the environment. In the highly dissipative regime, the quantum Zeno effect was seen, where strong environmental noise actually froze the dynamics of the system.
Together, these results show that our model is not only theoretically sound, but also practically realizable on today's NISQ hardware (noisy intermediate scale quantum)~\cite{Preskill2018NISQ,Kandala2017VQE,Temme2017Mitigation}.

To demonstrate the framework, we simulate two distinct types of noise: amplitude damping \cite{ Garraway1997, breuer}, which models energy loss and is non-unital; and random telegraph noise (RTN) \cite{banerjee2018open,benedetti,rice1992stochastic}, which is unital and represents phase fluctuations. 
In both cases, we show that the supervised ML model can accurately predict the system observable and that our revival-based metric successfully identifies and quantifies non-Markovianity.

\begin{figure}[h]
    \centering
    \includegraphics[width=1.0\linewidth]{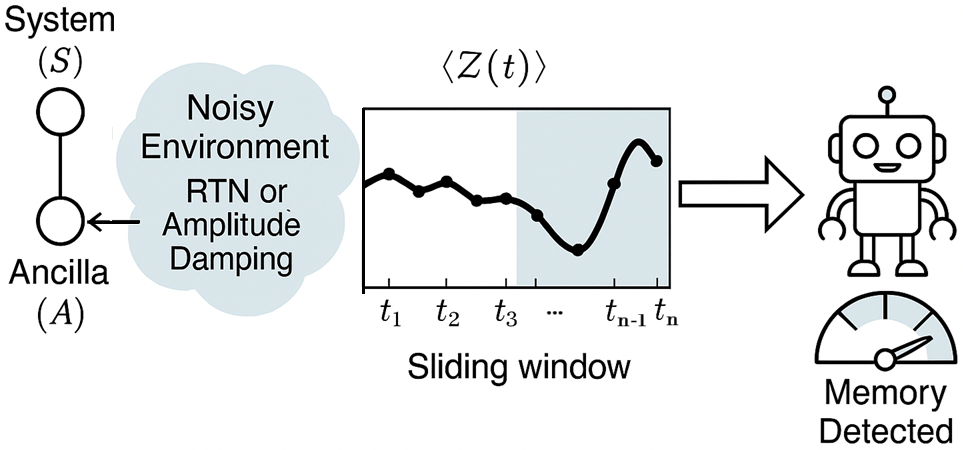}
    \caption{Schematic of the system-ancilla-environment setup. The system qubit (S) is coherently coupled to the ancilla qubit (A), which interacts with the environment (E). A supervised machine learning model observes the ancilla’s behavior to predict the system’s dynamics and detect non-Markovian memory effects.}
    \label{fig:fig1}
\end{figure}

Figure~\ref{fig:fig1} provides a visual overview of the setup. The system qubit (S) interacts coherently with the ancilla qubit (A), while only the ancilla is exposed to environmental noise. The supervised ML model monitors the ancilla's local observable $\langle Z_A(t) \rangle$ and uses it to infer $\langle Z_S(t) \rangle$, without access to the full state or environment.

The paper is organized as follows. Section~\ref{sec:sec2} introduces the system--ancilla--environment model and the noise channels considered. 
Section~\ref{sec:sec3} develops the open-system dynamics using time-local master equations for amplitude damping and RTN-induced dephasing. 
Section~\ref{sec:sec4} outlines the supervised learning framework, including dataset construction from ancilla measurements and the neural-network predictor for system observables. 
Section~\ref{sec:sec5} presents the second objective of this work: a bounded revival-based metric for detecting and quantifying non-Markovian memory effects. 
Section~\ref{sec:sec6} reports numerical simulations and compares the behavior under different noise regimes. 
Section~\ref{sec:sec7} discusses the relation of our approach to established non-Markovianity measures, in particular the BLP criterion. 
Finally, Section~\ref{sec:sec8} concludes with a summary of results and potential directions for future research.

\section{System--Ancilla--Environment Architecture and Noise Models} \label{sec:sec2}

We consider a minimal open quantum system consisting of two interacting qubits: a \textit{system qubit} ($S$), whose dynamics we aim to predict and analyze, and an \textit{ancilla qubit} ($A$), which is directly coupled to an external noisy environment. The ancilla acts as a probe through which environmental memory affects the system, as shown in Fig.~\ref{fig:fig1}.

The two qubits are coherently coupled via a symmetric XY interaction described by the Hamiltonian:
\begin{equation}
H_{(SA)} = g \left( X_{(S)} \otimes X_{(A)} + Y_{(S)} \otimes Y_{(A)} \right),
\label{eq:eq1}
\end{equation}
where \(X\) and \(Y\) are Pauli matrices and \(g\) is the coupling strength, which we set to 1 in dimensionless units. This interaction allows for unitary evolution between the qubits, enabling entanglement and information exchange.

The XY interaction in Eq.~\ref{eq:eq1} is deliberately chosen because it satisfies two key requirements for our framework. First, it entangles the system and ancilla qubits, which is essential for transferring the environmental influence to the system qubit through the ancilla. Second, it ensures that the time evolution of the relevant observables, such as \( \langle Z_{(S)}(t) \rangle \) and \( \langle Z_{(A)}(t) \rangle \), is non-trivial. This non-trivial evolution arises because the XY Hamiltonian does not commute with the \( Z \) operators of the system and ancilla. In contrast, if the interaction Hamiltonian were chosen to commute with \( Z_{(S)} \) or \( Z_{(A)} \) - for example, \( H_{(SA)} \propto Z_{(S)} \otimes Z_{(A)} \) - then the expectation values \( \langle Z_{(S)}(t) \rangle \) and \( \langle Z_{(A)}(t) \rangle \) would remain constant in time, completely masking any signatures of memory effects. Thus, the XY coupling plays a central role in making revival behavior detectable and experimentally accessible in our method.

To model the open system dynamics, we apply time-dependent noise to the ancilla only. We investigate two representative types of noise. First, we apply non-unital amplitude damping noise~\cite{ Garraway1997, breuer}, which causes energy dissipation from the excited state  to the ground state. The noise is implemented via a time-local Lindblad master equation with a time-dependent decay rate~\cite{vacchini2010exact, breuer}. This kind of noise not only introduces decoherence but also alters the populations of the ancilla, making it particularly suitable for studying energy-based memory effects. Second, we study a unital dephasing channel modeled by random telegraph noise (RTN)~\cite{Paladino2014,Cresser,PhysRevA.99.042128,sbrtn2} that lead to pure dephasing, causing coherence loss without population transfer. The RTN process is defined by an exponentially decaying autocorrelation function.

\section{Open-System Dynamics: Time-Local Master Equations for Amplitude Damping and RTN Dephasing}\label{sec:sec3}

    To simulate realistic open quantum dynamics, we model the interaction of the ancilla qubit with its environment using two representative noise channels. These channels determine how the environment induces decoherence, which is then transmitted to the system qubit through their coherent coupling with the ancilla qubit. The resulting dynamics generate observables such as \( \langle Z_{(S)}(t) \rangle \) and \( \langle Z_{(A)}(t) \rangle \), which together form the dataset used to train and test the supervised ML framework described in Section~\ref{sec:sec4}, as well as to implement the revival-based non-Markovianity detection method detailed in Section~\ref{sec:sec5}.

We describe both noise models using the time-local Lindblad master equation~\cite{lindblad,gorini}:
\begin{align}
\frac{d\rho_{(SA)}(t)}{dt}
&= -i[H^{(I)}_{(SA)}(t), \rho_{(SA)}(t)] \nonumber \\
&\quad + \sum_k \Big( L_k(t)\,\rho_{(SA)}(t)\,L_k^\dagger(t) \nonumber \\
&\quad
- \tfrac{1}{2}\{L_k^\dagger(t)L_k(t), \rho_{(SA)}(t)\} \Big).
\label{eq:eq2}
\end{align}

We work in the interaction picture with respect to $H_0=H_S+H_A+H_B$~\cite{breuer2002theory,Genes2019QPLMI}, where
$H_S=\tfrac{\omega_S}{2}Z_{(S)}$, $H_A=\tfrac{\omega_A}{2}Z_{(A)}$, and
$H_B=\sum_\ell \omega_\ell\, b_\ell^\dagger b_\ell$ are the free Hamiltonians of the system qubit, the ancilla qubit, and the bath modes, with $\omega_S,\omega_A,\omega_\ell$ their transition frequencies. The bath acts only on the ancilla, so $
L_k^{(I)}(t)=I_S\otimes L_k^{A,(I)}(t)$. Non-Markovian memory is captured by the explicit time dependence of the generator in the time-local form. The coherent system–ancilla coupling in Eq.~(\ref{eq:eq1}) is time independent. 
At resonance $(\omega_S=\omega_A)$, the interaction picture transformation 
$H_{SA}^{(I)}(t)=U_0^\dagger(t)H_{SA}U_0(t)$, where $U_{0} (t) = e^{-iH_{0}t}$, leaves the $XY$ coupling invariant 
(the $X$–$Y$ rotations cancel),  hence $H_{SA}^{(I)}(t)=H_{SA}$; thus Eq.~(\ref{eq:eq1}) holds in either picture~\cite[App.~C]{Genes2019QPLMI}, see also~\cite[Sec.~3.3]{breuer2002theory}.

We study two distinct types of noise: a non-unital amplitude damping channel that induces energy relaxation, and a unital dephasing channel induced by random telegraph noise (RTN), which preserves populations but degrades coherence. Each noise model leads to a different observable behavior in the system and ancilla, and ultimately affects the learning and detection outcomes.

\subsection{Non-Unital Amplitude Damping Channel}
Amplitude damping models the irreversible relaxation of a qubit’s excitation into the environment (e.g., spontaneous emission)~\cite{Garraway1997,breuer} and is well studied in the Markovian regime as well \cite{SGAD,OmkarSingleQubit}. We adopt the basis
$|0\rangle=\begin{pmatrix}1\\[2pt]0\end{pmatrix}$ (excited) and
$|1\rangle=\begin{pmatrix}0\\[2pt]1\end{pmatrix}$ (ground), whence $Z|0\rangle=+|0\rangle$, $Z|1\rangle=-|1\rangle$, and $\langle Z\rangle$ decays from $+1$ to $-1$. To write the energy–lowering jump with $\sigma_{-}$, we define on the ancilla
$\sigma_-^{A}:=|1\rangle\!\langle 0|$ (maps $|0\rangle\!\to|1\rangle$) and
$\sigma_+^{A}:=|0\rangle\!\langle 1|$.

With system–ancilla ordering $S\otimes A$, the ancilla-only collapse operator is

\begin{equation}
  L(t)=\sqrt{\gamma(t)}\,(I_S\otimes\sigma_-^{A})
  \label{eq:eq3}
\end{equation}
 and the Lindblad master equation is
\begin{align}
\frac{d\rho_{(SA)}}{dt}
&= -i\,[H_{(SA)},\,\rho_{(SA)}] \nonumber\\
& + \gamma(t)\Big[
(I_S\!\otimes\!\sigma_-^{A})\,\rho_{(SA)}\,(I_S\!\otimes\!\sigma_+^{A}) \nonumber\\
&- \tfrac{1}{2}\big\{\,I_S\!\otimes\!(\sigma_+^{A}\sigma_-^{A}),\,\rho_{(SA)}\big\}
\Big]
\label{eq:eq4}
\end{align}
which transfers population $|0\rangle\!\to|1\rangle$ on the ancilla, driving it toward $|1\rangle\!\langle1|$.

\begin{figure}[h!]
    \centering
    \includegraphics[width=1.0\linewidth]{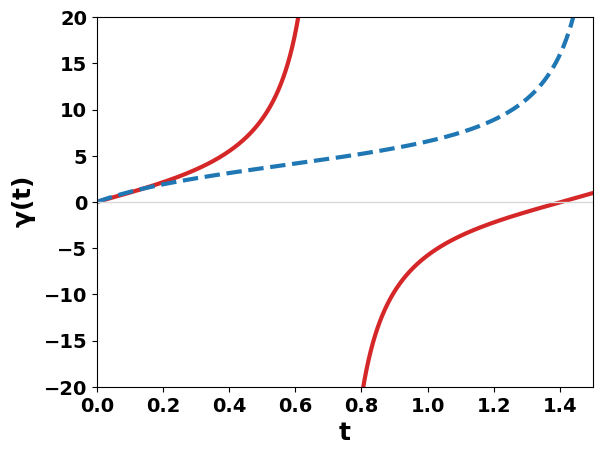}
\caption{Time-dependent decay rate \( \gamma(t) \) for the amplitude-damping model(Lorentzian reservoir). Two parameter sets are shown:
 red solid, non-Markovian, \( b=0.05,\ \lambda=10.0 \) (here \( b^{2}-2\lambda<0 \)), which produces temporary negativity; blue dashed, Markovian, \( b=5.0,\ \lambda=1.0 \) (here \( b^{2}-2\lambda>0 \)), for which \( \gamma(t)\ge 0 \) and the relaxation is monotonic.}
    
    \label{fig:fig2}
\end{figure}

The decay rate \( \gamma(t) \) is derived from a Lorentzian spectral density \cite{breuer2002theory,maniscalco} and is given by:
\begin{equation}
\gamma(t) = 1 - |G(t)|^2,
\label{eq:eq5}
\end{equation}
where
\begin{equation}
G(t) = e^{-bt/2} \left[ \cosh\left( \frac{dt}{2} \right) + \frac{b}{d} \sinh\left( \frac{dt}{2} \right) \right].
\label{eq:eq6}
\end{equation}
The decay rate is governed by \(d=\sqrt{b^{2}-2\lambda}\), where \(\lambda\) is the coupling strength and \(b\) is the spectral width of the reservoir, and the reservoir correlation time is $\tau_{B}=1/b$~\cite{haikka2010}. When \(b^{2} < 2\lambda\) (so $d$ is imaginary), \(\gamma(t)\) becomes temporarily negative, as shown in Fig.~\ref{fig:fig2}, signaling information backflow and hence non-Markovian dynamics. In contrast, for \(b^{2} >  2\lambda\) ( \(d\) is real), \(\gamma(t)\ge 0\) for all \(t\), corresponding to a monotonic (Markovian) relaxation.~\cite{PhysRevA.110.052209}. The reservoir correlation time is 

\subsection{Unital Dephasing Channel Induced by Random Telegraph Noise (RTN)}

To explore unital noise, we consider random telegraph noise (RTN)~\cite{Paladino2014,Bergli2007,PhysRevA.99.042128}, which causes phase damping but no energy loss. This noise model is classified as \textit{unital} because it preserves the maximally mixed state and affects only the off-diagonal elements of the density matrix. For example, an ancilla in the superposition state \( \ket{+} = (\ket{0} + \ket{1})/\sqrt{2} \) loses coherence over time but retains its populations.

RTN is modeled as a classical, zero-mean stochastic process $\xi(t)$ with autocorrelation
\begin{equation}
\langle \xi(t)\,\xi(s) \rangle = v^2\, e^{-\kappa |t-s|},
\label{eq:eq7}
\end{equation}
where $v$ is the noise amplitude and $\kappa$ is the correlation-decay rate (correlation time $\tau_B = 1/\kappa$) \cite{Bergli2007,sbrtn1,sbrtn2}.

The master equation under RTN, obtained from the corresponding Kraus operators \cite{Cresser,sbrtn1}, takes the form:
\begin{align}
\frac{d\rho_{(SA)}(t)}{dt}
&= -i[H_{(SA)}(t), \rho_{(SA)}(t)] \nonumber \\
&\quad + \Gamma(t)\Big[(I_{(S)} \otimes Z_{(A)})\,\rho_{(SA)}(t)\,(I_{(S)} \otimes Z_{(A)}) \nonumber \\
&\quad - \rho_{(SA)}(t)\Big],
\label{eq:eq10}
\end{align}

with time-dependent dephasing rate~\cite{breuer,breuer2002theory,Cresser}:
\begin{equation}
\Gamma(t) = -\frac{\dot{\Lambda}(t)}{2\,\Lambda(t)}.
\label{eq:eq11}
\end{equation}

Here, 

\begin{equation}
\Lambda(t) = e^{-\kappa t}\!\left[\cos\!\big(\chi\,\kappa t\big) + \frac{\sin\!\big(\chi\,\kappa t\big)}{\chi}\right],
\label{eq:eq8}
\end{equation}
with
\begin{equation}
\chi = \sqrt{\left(\frac{2v}{\kappa}\right)^{\!2} - 1 }.
\label{eq:chi}
\end{equation}

In our QuTiP simulations~\cite{qutip1,qutip2}, the decoherence is applied to the ancilla via the collapse operator
\begin{equation}
L(t) = \sqrt{\Gamma(t)}\,\big(I_{(S)} \otimes Z_{(A)}\big).
\label{eq:eq12}
\end{equation}

\begin{figure}[h!]
    \centering
    \includegraphics[width=1.0\linewidth]{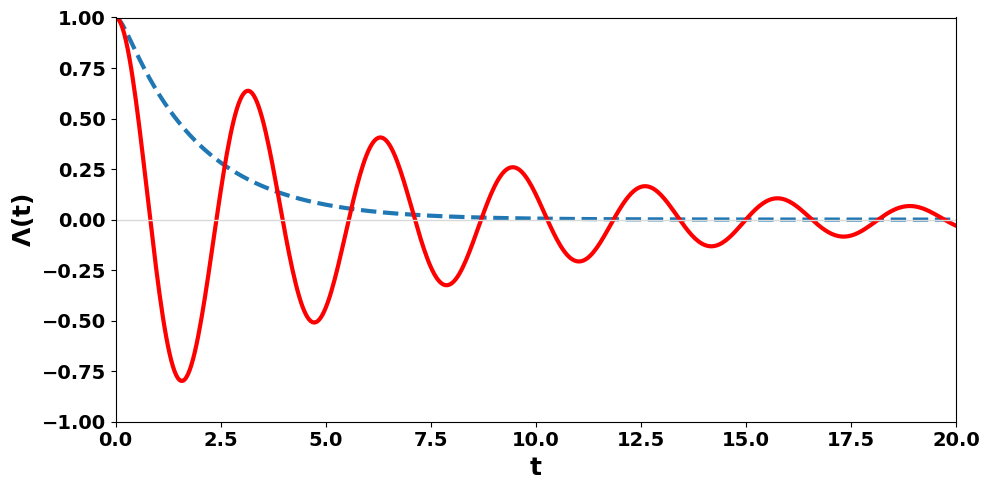}
\caption{Decoherence function $\Lambda(t)$ under RTN dephasing applied to the \emph{ancilla only} (collapse $I_S\!\otimes\! Z_A$). We use $r=v/\kappa$. 
{Blue dashed (Markovian):} $(v,\kappa)=(1,4)$, so $r=\tfrac{1}{4}=0.25<0.5$ (monotonic decay). {Red solid (non-Markovian):} $(v,\kappa)=(1,\,1/7)\approx(1,0.143)$, so $r=7.0>0.5$ (damped oscillations).}

    \label{fig:fig3}
\end{figure}

Hence, only the ancilla experiences direct dephasing, while the system responds indirectly through the shared XY coupling. The character of the dynamics depends on the ratio \(r := v/\kappa\). When \(r < 0.5\), \(\Lambda(t)\) decays smoothly, indicating Markovian behavior. When \(r > 0.5\), \(\Lambda(t)\) exhibits oscillations, revealing memory effects and non-Markovianity (since \(\chi=\sqrt{(2r)^2-1}\) becomes real). Figure~\ref{fig:fig3} shows these two regimes clearly.

\section{Supervised Learning Framework for Predicting System Observables}
\label{sec:sec4}

The goal of this section is to describe how we generate the quantum data and then train a supervised ML model to predict the system qubit’s observable \( \langle Z_{(S)}(t) \rangle \) using only recent measurements of a coupled ancilla qubit. To create realistic training data, we first simulate the open system dynamics by numerically solving the Lindblad master equation (Eq.~\ref{eq:eq2}) for the system-ancilla-environment setup described in Section~\ref{sec:sec3}. These simulations are carried out using the open-source quantum simulation library QuTiP~\cite{qutip1,qutip2}. From these simulations, we extract two quantities at each time step: the local ancilla measurement record and the true system observable \( Z_{(S)}(t) \). The true value of the system acts as the output label that the supervised ML model aims to predict. Following the standard supervised learning paradigm~\cite{Goodfellow2016}, the input features are constructed as a sliding window of the past ancilla measurements, while the output labels are the corresponding true observables of the system obtained from the simulation. The complete dataset of inputs and outputs is organized as shown in Table~\ref{tab:table1}. This dataset is then used to train the model to forecast future system dynamics using only partial, experimentally realistic ancilla data.

At each time step \(t_i\), we record two observables: the ancilla measurement \(Z_{(A)}(t_i)\), which is experimentally accessible, and the system observable \(Z_{(S)}(t_i)\), which is computed by solving the Lindblad master equation (see Section~\ref{sec:sec3}). The true system observable \(Z_{(S)}(t_i)\) acts as the \textit{label} for supervised learning: it provides the correct target the model must learn to predict. The input feature vector is constructed using a sliding window of the five most recent ancilla measurements: $
\mathbf{x}_i=\big[Z_{(A)}(t_{i-5}),\,Z_{(A)}(t_{i-4}),\,Z_{(A)}(t_{i-3}),\,Z_{(A)}(t_{i-2}),\,Z_{(A)}(t_{i-1})\big]$.
Since each input uses five prior ancilla measurements, the dataset begins at time step \(t_5\). Each row is then an input–output pair \((\mathbf{x}_i, y_i)\), where \(\mathbf{x}_i\) is the ancilla window and \(y_i = Z_{(S)}(t_i)\) is the true system observable at time \(t_i\).
For later reference, we denote the model’s prediction by \(\hat{y}_i\).

\begin{table}[h!]
\centering
\renewcommand{\arraystretch}{1.1} 
\setlength{\tabcolsep}{5pt}       
\small                            
\begin{tabular}{cc}
\toprule
\textbf{Input $\mathbf{x}_i$ (Ancilla Window)} & \textbf{Target Label $y_i = Z_{(S)}(t_i)$} \\
\midrule
$[Z_{(A)}(t_0),\,\ldots,\, Z_{(A)}(t_4)]$ & $Z_{(S)}(t_5)$ \\
$[Z_{(A)}(t_1),\,\ldots,\, Z_{(A)}(t_5)]$ & $Z_{(S)}(t_6)$ \\
$[Z_{(A)}(t_2),\,\ldots,\, Z_{(A)}(t_6)]$ & $Z_{(S)}(t_7)$ \\
$\vdots$ & $\vdots$ \\
\bottomrule
\end{tabular}
\caption{Compact structure of the supervised learning dataset: each input is a window of past ancilla measurements, and each output is the true system observable at the next time step.}
\label{tab:table1}
\end{table}

During training, the goal of the supervised ML model is to learn how to map each input feature vector \( \mathbf{x}_i \) to a predicted output \( \hat{y}_i \) that accurately estimates the system observable at that time step. The output \( \hat{y}_i \) represents the model’s best prediction of the system’s Pauli-\(Z\) expectation value, inferred solely from the ancilla’s measurement window. By comparing this prediction with the known simulation value, \( y_i \) (the target label shown in Table~\ref{tab:table1}), the model iteratively adjusts its internal weights to minimize the difference. The remainder of this section explains the complete methodology and the model architecture in detail.

\begin{figure}[h]
    \centering
    \includegraphics[width=1.0\linewidth]{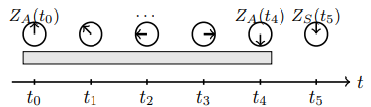}
    \caption{Illustration of the sliding window used during training. At each time step \( t_i \), the supervised ML model receives the previous five ancilla measurements \( [Z_{(A)}(t_{i-5}), \ldots, Z_{(A)}(t_{i-1})] \) as input and learns to predict the system observable \( Z_{(S)}(t_i) \), which serves as the target label. This process generates the input–output pairs summarized in Table~\ref{tab:table1}.}
    \label{fig:fig4}
\end{figure}

The overall data preparation process is shown schematically in Fig.~\ref{fig:fig4}. As the sliding window moves forward through the simulated ancilla data, it generates a sequence of input–output pairs $(\mathbf{x}_i, y_i)$, where each input window provides the supervised ML model with local temporal information needed to predict the next system observable. Together, these pairs form the complete dataset for training and testing the neural network, structured exactly as summarized in Table~\ref{tab:table1}.

To ensure reliable and generalizable predictions, we split the dataset from Table~\ref{tab:table1} (constructed via the sliding window in Fig.~\ref{fig:fig4}) into a time-ordered 50\%/50\% partition: the first half is used for training and the second half for testing, with no shuffling to avoid temporal leakage. Our predictor is a \textit{feedforward neural network}~\cite{Goodfellow2016}, a standard supervised architecture for time-series regression. Although other supervised models (e.g., decision trees~\cite{Quinlan1996} or support vector machines~\cite{Cortes1995}) could be employed, we adopt this simple feedforward design for clarity and scalability.

The network comprises an input layer, two hidden layers, and a single-output layer (Fig.~\ref{fig:fig5}). The hidden layers introduce nonlinear transformations that enable learning of structured temporal patterns in the ancilla measurements—patterns not captured by a purely linear map. This allows the model to map a short sequence of past ancilla measurements to the corresponding future system observable.

Training minimizes the mean-squared error (MSE) between the prediction \(\hat{y}_i\) and the ground-truth label \(y_i\):
\begin{equation}
\text{MSE} = \frac{1}{N_{\text{train}}} \sum_{i=1}^{N_{\text{train}}} \big( y_i - \hat{y}_i \big)^2,
\label{eq:eq13}
\end{equation}
where \(N_{\text{train}}\) is the number of training samples. Model parameters are optimized with the \textit{Adam} algorithm~\cite{Kingma2014adam}. 

\subsection{Neural Network Architecture}

\begin{figure*}[t]
\centering
\includegraphics[width=0.95\linewidth]{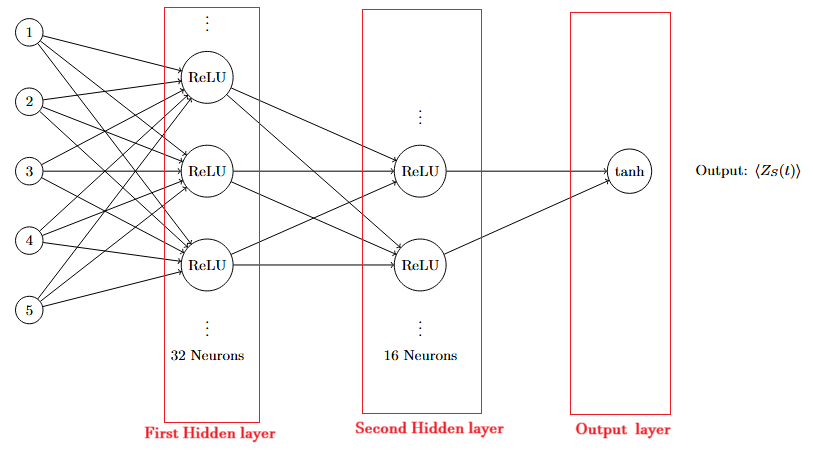}
\caption{Neural network used for system prediction. The input layer receives 5 recent ancilla values. Two hidden layers (32 and 16 neurons) use ReLU activation, and the output neuron applies a \texttt{tanh} activation to produce \( \hat{y}_i \in [-1, 1] \).}
\label{fig:fig5}
\end{figure*}

The neural network in our framework is designed with three main parts: an input layer, two hidden layers, and an output layer. Together, these layers enable the model to learn the complex, non-linear relationship between a short sequence of recent ancilla measurements and the future system observable. The hidden layers, in particular, introduce non-linear transformations that allow the network to capture patterns that a simple linear mapping could not detect. This design forms the foundation for accurately predicting the system’s dynamics using only local ancilla data. The detailed structure and function of each layer are described in this section. Figure~\ref{fig:fig5} shows an overview of the complete network. Each part of the neural network is fully connected, meaning that every neuron in one layer is connected to every neuron in the next layer. In our design:

\begin{itemize}
    \item \textbf{First hidden layer:} This layer consists of 32 neurons. Each neuron receives all five input features from the sliding ancilla window. Each neuron computes a weighted sum of these inputs, adds a bias term, and passes the result through a Rectified Linear Unit (ReLU) activation function~\cite{nair2010relu} to introduce non-linearity:
    \begin{equation}
        h^{(1)}_j = \text{ReLU} \Big( \sum_{k=1}^{5} w^{(1)}_{jk} x_k + b^{(1)}_j \Big),
        \label{eq:eq14}
    \end{equation}
    where  \( h^{(1)}_j \) is the output of the $j$th neuron in the first hidden layer, \( x_k \) is the $k$th input feature (ancilla measurement),  \( w^{(1)}_{jk} \) is the weight connecting input feature $k$ to neuron $j$, and  \( b^{(1)}_j \) is the bias for neuron $j$. The ReLU  activation function is defined as:
\[
\text{ReLU}(z) = 
\begin{cases}
z, & z > 0 \\ 
0, & z \leq 0.
\end{cases}
\]
This means that if the neuron’s total input $z$ (the weighted sum plus bias) is positive, the neuron outputs that value and is said to \textit{fire}. If the input is zero or negative, the neuron outputs zero and does not activate. This simple non-linear rule enables the network to pass forward only meaningful positive signals, helping it learn complex patterns while keeping the computations efficient.

    \item \textbf{Second hidden layer:} This layer has 16 neurons. Each neuron here is fully connected to all 32 outputs from the first hidden layer. Each neuron again computes a weighted sum, adds a bias, and applies the same ReLU activation:
    \begin{equation}
        h^{(2)}_j = \text{ReLU} \Big( \sum_{k=1}^{32} w^{(2)}_{jk} h^{(1)}_k + b^{(2)}_j \Big),
        \label{eq:eq15}
    \end{equation}
    where \( h^{(2)}_j \) is the output of the $j$th neuron in the second hidden layer,  \( h^{(1)}_k \) is the output from the $k$th neuron of the first hidden layer,  \( w^{(2)}_{jk} \) is the weight connecting neuron $k$ to neuron $j$, and \( b^{(2)}_j \) is the bias for neuron $j$.

    \item \textbf{Output layer:} The output layer contains a single neuron that takes input from all 16 outputs of the second hidden layer. It combines these using learned weights and a bias, then applies a hyperbolic tangent (\(\tanh\)) activation function to ensure the output stays within the physical range \([-1, 1]\):
    \begin{equation}
        \hat{y}_i = \tanh\Big( \sum_{j=1}^{16} w^{(3)}_j h^{(2)}_j + b^{(3)} \Big),
        \label{eq:eq16}
    \end{equation}
 where, \( \hat{y}_i \) is the predicted system Pauli-$Z$ expectation at time step $t_i$,  \( w^{(3)}_j \) is the weight connecting the $j$th neuron of the second hidden layer to the output,  \( b^{(3)} \) is the output neuron’s bias, and  \( h^{(2)}_j \) is the output of the $j$th neuron in the second hidden layer.
    
\end{itemize}

Each neuron in every layer thus performs a simple computation: it sums its weighted inputs, adds a bias, and applies a non-linear activation function to produce its output, which then feeds forward to the next layer.

As described above, the supervised ML model is trained by minimizing the mean squared error (MSE) between its predicted output and the true system observable, as defined in Eq.~\ref{eq:eq13}. The Adam optimizer is used to adjust the network’s weights during this process. Once training is finished, the model can make predictions on new data that it has not seen before. In other words, the trained network can estimate the system’s observable using only a short sequence of recent ancilla measurements, without needing full access to the system’s quantum state. This makes the method practical and suitable for real experiments.

All the procedures described above, including the construction of the input–output pairs $(\mathbf{x}_i, y_i)$, the architecture of the neural network, and the optimization process, are part of the training phase. During this phase, the supervised ML model is exposed to {50\%} of the dataset, where both the input $\mathbf{x}_i$ and the true output $y_i$ (the label) are known. The network learns to associate patterns in $\mathbf{x}_i$ with the correct label $y_i$ by adjusting its internal weights to minimize the prediction error.

Once the model has acquired sufficient experience from these training examples, we transition to the prediction phase. In this phase, the remaining {50\%} of the dataset is used for testing. Here, the model receives only the input windows $\mathbf{x}_i$ without access to the true labels $y_i$. Unlike the training phase, no further weight updates are performed. Instead, the model relies solely on the weights and patterns it learned during training to predict the system observable $\hat{y}_i$. This prediction phase demonstrates the model’s ability to generalize and infer quantum dynamics independently, using only ancilla measurements.

\section{Revival-Based Metric for Detecting and Quantifying Non-Markovian Memory} \label{sec:sec5}

\begin{figure}[h]
    \centering
    \includegraphics[width=1.0\linewidth]{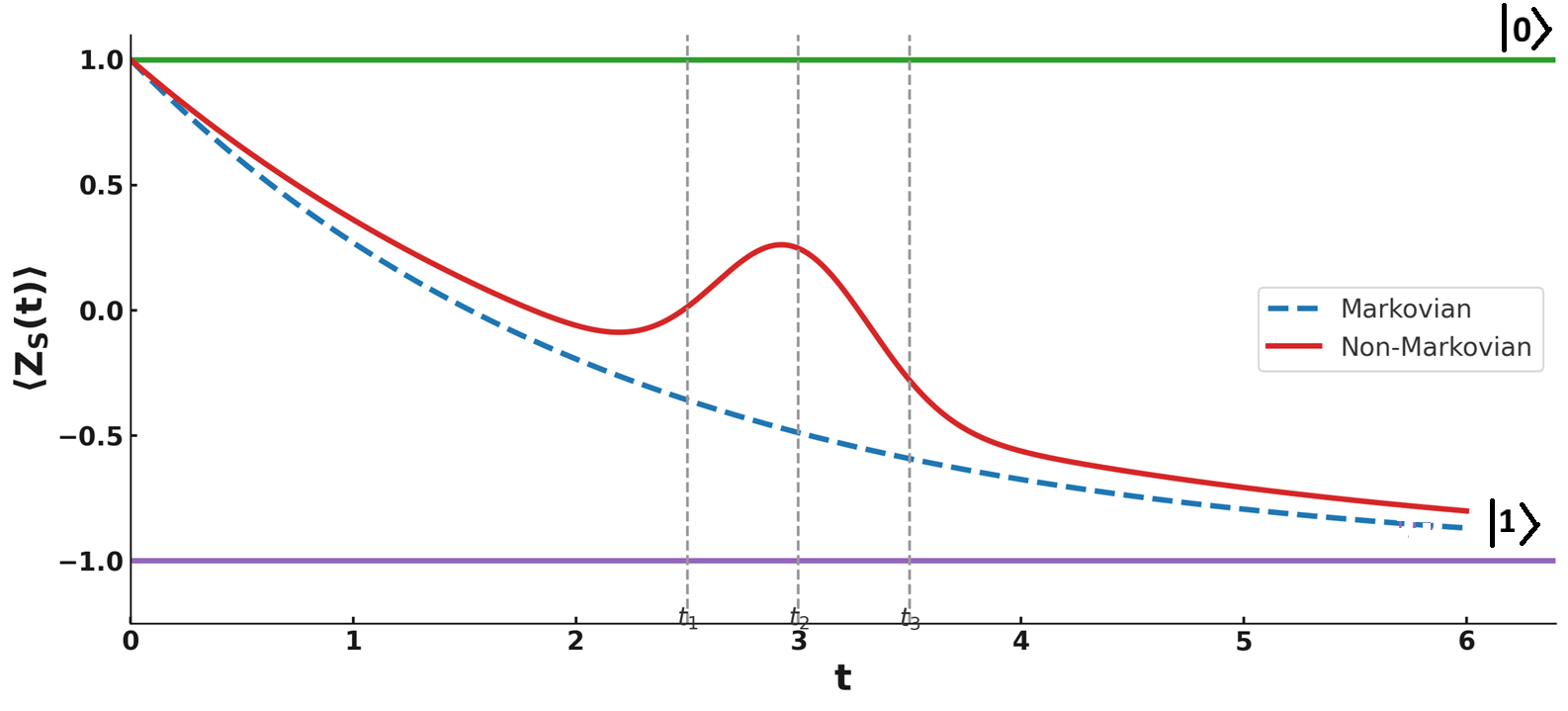}
       \caption{Schematic of the system observable \(\langle Z_{S}(t)\rangle\). We take \(|0\rangle\) (excited state) at \(+1\) and \(|1\rangle\) (ground state) at \(-1\). The solid red curve shows non-Markovian dynamics with a clear revival between \(t_1\) and \(t_3\) (local maximum at \(t_2\)), indicating temporary information backflow, while the dashed blue curve depicts a monotonic Markovian relaxation toward the ground state.}
    \label{fig:fig6}
\end{figure}

In this section, we present our second main goal: detecting and quantifying non-Markovian memory effects in open quantum systems. Unlike traditional methods that require full access to the quantum state, our approach works entirely with observable quantities. Specifically, we use the predicted values of the system observable $\hat{y} = Z_{(S)}(t)$ obtained from the trained supervised ML model described in Section~\ref{sec:sec4}. These predictions rely solely on local ancilla measurements, making our method practical and experimentally feasible.
\noindent
The key idea is that non-Markovian dynamics manifest as temporary reversals in the system's evolution. Such reversals, commonly known as \textit{revivals}, indicate the backflow of information from the environment to the system. In Markovian systems, this kind of behavior is absent: the system observables change smoothly in one direction, typically relaxing monotonically toward equilibrium over time. A \textit{revival} refers to a temporary reversal in the evolution of the system observable \( \langle Z_{(S)}(t) \rangle \). Depending on the system's initial state and the type of noise, this can appear as either a local increase or a local decrease. For example, if the system starts in the excited state \( \ket{0} \) (with \( \langle Z_{(S)}(0) \rangle = +1 \)), a revival might show up as a temporary upward rise in \( \langle Z_{(S)}(t) \rangle \), signaling a partial recovery of excitation due to memory effects. Likewise, if the system starts near the ground state, a revival could appear as a temporary dip before continuing toward its steady state. 
\noindent
These non-monotonic behaviors are hallmarks of \textit{non-Markovian} dynamics, where information flows back from the environment into the system. In contrast, \textit{Markovian} dynamics result in a smooth, monotonic relaxation without revivals. Figure~\ref{fig:fig6} illustrates this concept, the dashed line shows a typical Markovian evolution, where the system observable steadily moves toward equilibrium without interruption. The solid line represents non-Markovian dynamics, showing a revival between times \( t_1 \) and \( t_3 \), where the observable rises and then falls again. This behavior indicates the presence of environmental memory effects.

\paragraph*{Revival detection.}
In our convention, the \emph{predicted} system trajectory \(\widehat{Z}_{(S)}(t)\) typically drifts downward toward \(-1\); a \emph{revival} is a short upward run in this predicted trajectory (i.e., a contiguous segment with \(\Delta \hat{y}_i=\hat{y}_i-\hat{y}_{i-1}>\varepsilon\)).

Let $\hat y_i=\langle Z_S(t_i)\rangle$ and define the step
\[
\Delta \hat y_i=\hat y_i-\hat y_{i-1}.
\]
With a small threshold $\varepsilon>0$ (we use $\varepsilon=0.015$):
\begin{itemize}
  \item \textbf{Start ($t_1$):} the first index with $\Delta\hat y_i>\varepsilon$.
  \item \textbf{Between $t_1$ and $t_2$:} the curve keeps rising, so consecutive steps satisfy $\Delta\hat y_i>0$; the \textbf{peak} $t_2$ is the last point in this rising run.
\end{itemize}
A clear \emph{hallmark of a revival} is that immediately after the peak, $\Delta\hat y$ switches from positive to non-positive (becomes $\le 0$).

To \textit{quantify} the degree of non-Markovianity, we define a revival-based non-Markovianity measure \( \mathcal{M}_{\text{rev}} \) using the predicted values of the system observable \( {Z}_{(S)}(t_i) \) produced by the trained RL model. A revival is detected whenever the change in the predicted observable exceeds a small positive threshold \( \varepsilon \). The measure is given by:

To quantify the degree of non-Markovian behavior in the system, we define a revival-based measure \( \mathcal{M}_{\text{rev}} \) that counts the number of revivals detected in the predicted dynamics of the system observable. A revival is identified as a significant change in the system’s behavior that indicates a temporary reversal, often interpreted as a sign of information flowing back from the environment to the system.
The measure is defined as:
\begin{equation}
\mathcal{M}_{\text{rev}} = \sum_{i=1}^{N_{\mathrm{eval}}} \Theta\!\left( \Delta {Z}_{(S)}(t_i) - \varepsilon \right),
\label{eq:eq19}
\end{equation}
where \( \Delta{Z}_{(S)}(t_i)=\hat{Z}_{(S)}(t_i)-\hat{Z}_{(S)}(t_{i-1}) \), \( \varepsilon \) is a small threshold value used to filter out minor fluctuations due to noise or numerical error, and \( N_{\mathrm{eval}} \) denotes the number of \emph{evaluated} time samples on which revivals are detected (in our experiments, the test subset).

The function \( \Theta(x) \) is the Heaviside step function~\cite{heaviside1892electrical}:
\begin{equation}
\Theta(x) =
\begin{cases}
1, & \text{if } x > 0, \\
0, & \text{otherwise}.
\end{cases}
\label{eq:eq20}
\end{equation}

\vspace{1em}
\noindent
\textbf{Example.}
Let the predicted observable over six times be
\[
{Z}_{(S)}=[0.90,\ 0.92,\ 0.94,\ 0.93,\ 0.95,\ 0.97],\qquad
\varepsilon=0.015.
\]
Use the \emph{forward} difference to detect upward steps:
\[
\Delta{Z}_{(S)}(t_i)={Z}_{(S)}(t_i)-{Z}_{(S)}(t_{i-1}).
\]
Then
\[
\begin{aligned}
\Delta{Z}_{(S)}(t_2)&=0.92-0.90=0.02,\\
\Delta{Z}_{(S)}(t_3)&=0.94-0.92=0.02,\\
\Delta{Z}_{(S)}(t_4)&=0.93-0.94=-0.01,\\
\Delta{Z}_{(S)}(t_5)&=0.95-0.93=0.02,\\
\Delta{Z}_{(S)}(t_6)&=0.97-0.95=0.02.
\end{aligned}
\]
Apply the revival test \(\Theta(\Delta{Z}_{(S)}(t_i)-\varepsilon)\):
\[
\begin{aligned}
&\Theta(0.02-0.015) = 1, \quad  \Theta(0.02-0.015) = 1, \\
& \Theta(-0.01-0.015) = 0, \quad \Theta(0.02-0.015) = 1, \\
&\Theta(0.02-0.015) = 1.
\end{aligned}
\]
Thus the revival measure is
\[
\mathcal{M}_{\mathrm{rev}}=1+1+0+1+1=4.
\]

We found four short “turn-backs” (revivals) in the predicted signal. A raw count increases simply because we observe more time points, so it cannot be fairly compared across runs of different lengths. For example, running the dynamics for a longer time usually yields more revivals simply because there are more opportunities to observe them; this does not indicate stronger non-Markovianity. Therefore, we introduce  a normalized score by dividing the revival count by the number of evaluated time points.

At each evaluated time point $t_i$ (the test samples), we check whether the signal
increases more than a small threshold $\varepsilon$ compared to the previous point:
$\Delta Z_{(S)}(t_i)= Z_{(S)}(t_i)- Z_{(S)}(t_{i-1})>\varepsilon$.
Each such point is counted as a \emph{revival-like upward move}. The Normalized metric is therefore
\begin{equation}
 \mathcal M_{\text{rev}}^{(\text{norm})}
= \frac{N_{\text{rev}}}{N_{\text{eval}}}
= \frac{1}{N_{\text{eval}}}\sum_{i=1}^{N_{\text{eval}}}
\Theta\!\big(\Delta\hat Z_{(S)}(t_i)-\varepsilon\big),
\label{eq:eq19}
\end{equation}

where $N_{\text{rev}}$ is the number of time points with an upward step $>\varepsilon$,
and $N_{\text{eval}}$ is the total number of evaluated time points (the test set).
This gives the \emph{fraction of evaluated samples} that exhibit a revival-like move,
so $0\le \mathcal M_{\text{rev}}^{(\text{norm})}\le 1$.

For example, if we observe  $4$ revivals in $200$ evaluated points, then $4/200=0.02$.
If we double the horizon and see $8$ revivals in $400$ points, we still get $8/400=0.02$.
The score stays the same, showing the dynamics did not change—we simply looked longer.

$\mathcal M_{\text{rev}}^{(\text{norm})}$ is the probability (per evaluated time step)
of observing a revival-like upward move above $\varepsilon$, making results comparable
across different durations and splits.

Crucially, this detection and quantification process uses only the predicted values \( {Z}_{(S)}(t_i) \), without requiring the true values of the system \( Z_{(S)}(t_i) \). This mimics realistic experimental scenarios where the system state is not directly accessible and measurements are inferred through ancillary observations and learned models. Despite this limited access, our revival-based method provides a robust and interpretable measure of non-Markovianity.

\section{Numerical Results and Discussion}\label{sec:sec6}

We evaluate our approach on the two noise models introduced earlier: (i) the non-unital amplitude-damping channel and (ii) the unital dephasing channel driven by random telegraph noise (RTN). For each model we explore both Markovian and non-Markovian parameter regimes and apply the supervised predictor together with the revival detector of Sec.~\ref{sec:sec2} (threshold $\varepsilon=0.015$).

For amplitude damping, with the excited/ground labeling fixed above, the baseline $\langle Z_S(t)\rangle$ drifts toward $-1$: it is monotonic in the Markovian case, whereas in the non-Markovian case it shows brief upward excursions (revivals). Our detector extracts the start and peak times $(t_1,t_2)$ of these excursions directly from $\langle Z_S(t)\rangle$ using only local, experimentally accessible data. For RTN dephasing, populations remain fixed while coherence decays; Markovian RTN yields a smooth loss of coherence, while non-Markovian RTN exhibits transient recoveries. The same revival criterion flags these coherence comebacks without requiring access to the environment.

Across both case studies, the model predicts near-term dynamics and flags non-Markovian features from local data alone, demonstrating that the same supervised pipeline transfers cleanly between non-unital and unital noise settings.

\subsection{Supervised ML Performance under Non-Unital Noise (Amplitude Damping)}

\begin{figure*}[t]
    \centering
    \subfloat[Markovian regime \label{fig:fig7a}]{
                \includegraphics[width=0.8\linewidth]{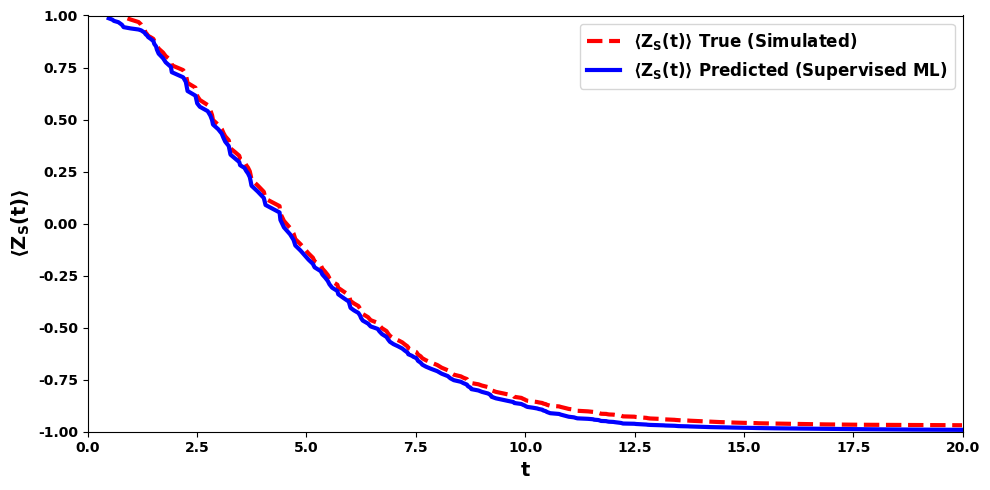}}
    
    \subfloat[Non-Markovian regime\label{fig:fig7b}]{
            \includegraphics[width=0.8\linewidth]{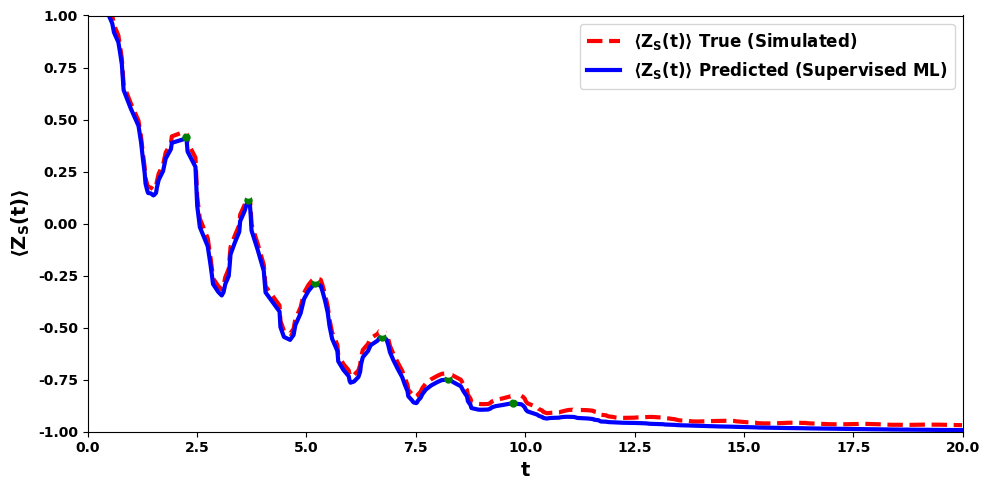}}
\caption{Amplitude–damping dynamics with the convention \(|0\rangle\) (excited) at \(+1\), \(|1\rangle\) (ground) at \(-1\). Dashed red: exact simulation; solid blue: supervised-ML prediction; and Green dots: Predicted revival peaks in $\langle  Z_S(t)\rangle $ caused by non-Markovian information/energy backflow (partial re-excitation) before the decay resumes. \textbf{(a)} Markovian regime, \(b=5.0,\ \lambda=1.0\): smooth monotonic relaxation with no revivals. \textbf{(b)} Non-Markovian regime, \(b=0.05,\ \lambda=10.0\): damped oscillations (revivals) before relaxation, indicating information backflow.}

    \label{fig:fig7}
\end{figure*}

To evaluate the supervised ML's ability to capture the effects of non-unital noise, we test its performance under both Markovian and non-Markovian regimes of amplitude damping. In each case, we compare the predicted observable \( \hat{y} = \langle Z_{(S)}(t) \rangle \) with the exact evolution \( y \) obtained by numerically solving the Lindblad master equation.

In this setup, both the system and ancilla qubits are initialized in the excited state \( \ket{0} \), which corresponds to an initial expectation value \( \langle Z_{(S)}(0) \rangle = +1 \). As the system evolves under amplitude damping noise, it steadily relaxes toward the ground state \( \ket{1} \), for which \( \langle Z_{(S)}(t) \rangle = -1 \). In the Markovian case, this evolution is smooth and irreversible, with no revivals, clearly indicating a lack of memory effects. The supervised ML  predicted trajectory closely matches the exact solution, demonstrating its ability to accurately capture the dynamics of amplitude damping without falsely identifying normal decay as non-Markovian behavior.

\paragraph*{Markovian Case.}  
Figure~\ref{fig:fig7a} shows the system’s behavior under amplitude damping noise in the Markovian regime. This is achieved by setting a large spectral width \( b = 5 \) and weak coupling \( \lambda = 1 \), which keeps the decoherence rate \( \gamma(t) \) smooth and positive, as defined in Eq.~\ref{eq:eq6}.  Both the system and ancilla qubits are initially prepared in the excited state \( \ket{0} \), resulting in an initial expectation value \( \langle Z_{(S)}(0) \rangle = +1 \). As the amplitude damping process unfolds, the system relaxes steadily toward the ground state \( \ket{1} \), causing the observable \( \langle Z_{(S)}(t) \rangle \) to decrease monotonically toward \( -1 \). This smooth, irreversible behavior is typical of Markovian dynamics, where no memory-induced revivals occur.  

In the figure, the dashed red curve shows the true system observable computed from the Lindblad master equation, while the solid blue curve shows the supervised ML model’s prediction. The close match between them demonstrates that the model accurately learns the true evolution and correctly identifies the absence of non-Markovian effects in this regime.
\paragraph*{Non-Markovian Case.}  
Figure~\ref{fig:fig7b} shows the system’s dynamics under amplitude damping in the non-Markovian regime, where the parameters are set to a narrow spectral width \( b = 0.05 \) and strong coupling \( \lambda = 10 \). This setup makes the environment highly correlated in time, so information that leaves the system can flow back, producing memory effects.  

Unlike the smooth, monotonic relaxation seen in the Markovian case, the system observable \( \langle Z_{(S)}(t) \rangle \) here exhibits clear \textit{revivals}: the trajectory temporarily reverses, showing upward oscillations during its overall relaxation toward the ground state. These oscillations are direct signatures of non-Markovian information backflow.  

In the plot, the dashed red curve shows the true observable from the numerical simulation, while the solid blue curve shows the prediction by the supervised ML model. The close match confirms that the model accurately captures the timing and strength of the revival events, using only local ancilla measurements. This result demonstrates the model’s ability to detect and reproduce non-Markovian memory behavior without requiring full access to the system’s state.

\subsection{Supervised ML Performance under Unital Noise (RTN Dephasing)} 
\label{sec:unital}

To test the generality of our supervised ML framework, we apply it to a pure dephasing noise model based on Random Telegraph Noise (RTN). Unlike amplitude damping, RTN affects only the system’s coherence without changing its populations. Despite this difference, the supervised ML model--trained solely on past ancilla measurements--can still learn to predict how the system observable \( \langle Z_{(S)}(t) \rangle \) evolves under dephasing. We initialize the joint state as $
\psi_0=\ket{+}_S\otimes\ket{0}_A,\quad
\ket{+}=\tfrac{1}{\sqrt{2}}(\ket{0}+\ket{1})$,
where, by our convention, the ancilla’s \emph{excited} state is \(\ket{0}_A\) (ground \(=\ket{1}_A\)).
RTN dephasing acts on the ancilla along \(Z_A\), leaving its populations fixed while randomizing its phase.
Through the system–ancilla \(XY\) coupling \(H_{SA}\) (which does not commute with \(Z_S\)), these phase kicks induce a nontrivial evolution of \(\langle Z_{(S)}(t)\rangle\).

\paragraph*{Markovian Regime.}

\begin{figure*}[t]
    \centering
    \subfloat[Markovian regime \label{fig:fig8a}]{
        \includegraphics[width=0.8\linewidth]{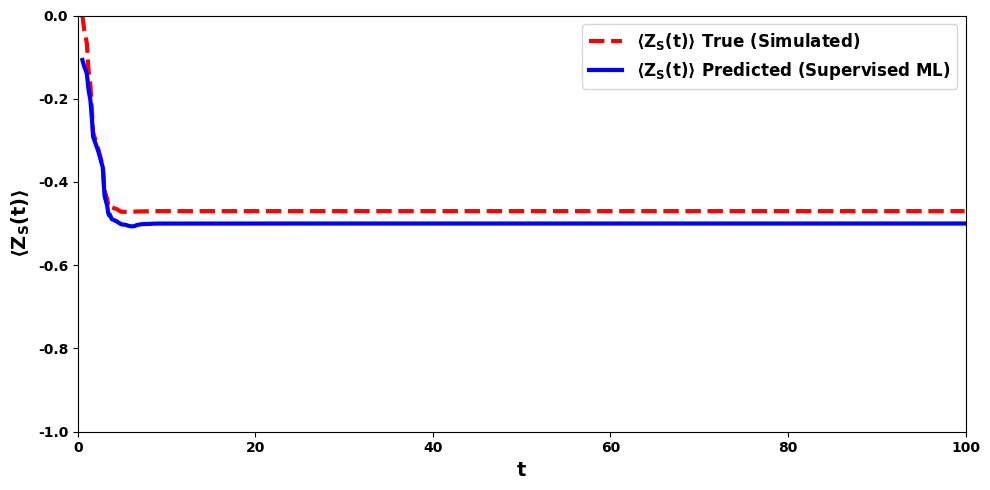}
    }
    \par\bigskip
    \subfloat[Non-Markovian regime \label{fig:fig8b}]{
        \includegraphics[width=0.8\linewidth]{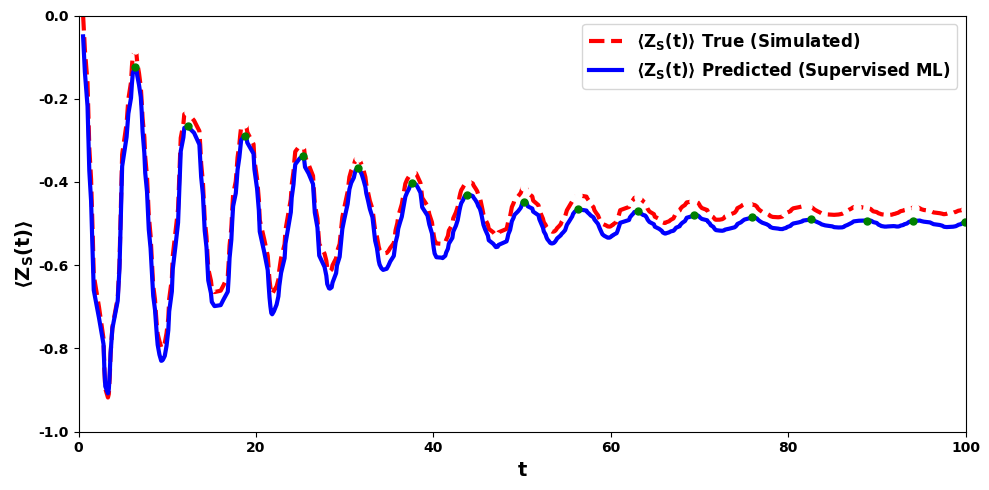}
    }
    \caption{System dynamics under RTN-induced pure dephasing noise (applied to the ancilla). The system starts in the superposition \(\ket{+}\). The dashed red line is the exact solution of the Lindblad master equation with RTN dephasing; the solid blue line is the supervised-ML prediction from local ancilla data ; and green dots are the  predicted revival peaks in $\langle Z_S(t)\rangle $ caused by non-Markovian information/coherence backflow under dephasing, before the decay resumes.(a) Markovian regime, \(r=v/\kappa<0.5\): smooth decay without revivals. (b) Non-Markovian regime, \(r>0.5\): damped oscillations (revivals) appear due to environmental memory.}

    \label{fig:fig8}
\end{figure*}

Figure~\ref{fig:fig8a} shows the system observable \( \langle Z_{(S)}(t) \rangle \) under RTN dephasing in the Markovian regime ($r=v/\kappa < 0.5$)
. In this case, the environment has no memory, so the coherence decays smoothly and monotonically. Starting from \( \langle Z_{(S)}(0) \rangle = 0 \), the observable gradually drops and approaches a steady value near \(-0.5\), reflecting the initial population imbalance. The dashed red curve shows the simulated solution from the Lindblad master equation, while the solid blue curve shows the supervised ML prediction based only on ancilla data. Their close match confirms that the model correctly reproduces the Markovian behavior without introducing false signs of memory effects.

\paragraph*{Non-Markovian case.}
As shown in Fig.~\ref{fig:fig8b}, in the RTN non-Markovian regime ($r=v/\kappa>0.5$) the environment retains memory, producing \emph{pronounced and long-lived} revivals in the system observable. The simulated reference (dashed red) and the supervised prediction from ancilla-only inputs (solid blue) agree closely, indicating that the model captures these non-Markovian features without access to the full state. Compared with the non-Markovian amplitude-damping case (Fig.~\ref{fig:fig7b}), RTN revivals occur more frequently and persist over a longer temporal window, consistent with the larger normalized revival score reported in Sec.~\ref{sec:sec6}.

\subsection{Non-Markovianity Quantification}
We quantify non-Markovianity with the \emph{normalized revival score} in Eq.~\eqref{eq:eq19}, which counts how many upward bumps (\(\Delta Z_i>\varepsilon\)) appear in \(\langle Z_{(S)}(t)\rangle\) and divides by the number of evaluated samples \(N_{\mathrm{eval}}\) (here \(N_{\mathrm{eval}}=500\)). For non-Markovian amplitude damping (non-unital) and non-Markovian RTN dephasing (unital) we obtain
\(N_{\mathrm{rev}}^{\mathrm{AD}}=7\) and \(N_{\mathrm{rev}}^{\mathrm{RTN}}=16\),
hence
\(\mathcal{M}_{\mathrm{rev}}^{\mathrm{AD\,(norm)}}=7/500=0.014\) and
\(\mathcal{M}_{\mathrm{rev}}^{\mathrm{RTN\,(norm)}}=16/500=0.032\).
\noindent
We kept the numerical settings the same for both runs (same time step, total time,  and threshold $\varepsilon$), so the scoring method is comparable. The normalized revival scores are $0.014$ (amplitude damping) and $0.032$ (RTN), is approximately  $2.3$ times higher for RTN. Because the initial states are not identical, this is an indicative (not exact) comparison; nevertheless, under these conditions RTN shows stronger non-Markovian behavior by our measure.

\section{Relation to the BLP Non-Markovianity Criterion}\label{sec:sec7}

The measure introduced here is \emph{operationally aligned} with the Breuer–Laine–Piilo (BLP) notion of non-Markovianity~\cite{breuer}.  
BLP flags memory effects when the trace distance between two system states becomes non-monotonic (information backflow), but it typically requires state tomography to track the full density matrices.

By contrast, we use a single, directly observable signal $\langle Z_{(S)}(t)\rangle$ and count its upward excursions (“revivals”).  
In our protocols (amplitude damping and RTN), these revivals occur precisely when BLP would register a non-monotonic change in distinguishability; hence, the \emph{events} we flag are the same phenomenon BLP targets, but \emph{without} tomography or process reconstruction.  
Thus, our method offers a lightweight, experimentally friendly proxy for BLP-style non-Markovianity.

\section{Conclusion}
\label{sec:sec8}

We developed a supervised machine-learning framework to detect and quantify non-Markovian memory in open quantum systems using only local ancilla measurements. A system qubit is coherently coupled to an ancilla via an XY interaction; the ancilla alone is exposed to noise and is measured. A central contribution is a bounded, revival-based metric \(\mathcal{M}_{\text{rev}}\) that quantifies memory by identifying non-monotonic features (revivals) in the \emph{predicted} system observable \(\langle Z_{(S)}(t)\rangle\), without state tomography or detailed bath knowledge. We demonstrated the approach on two contrasting channels: non-unital amplitude damping (energy relaxation) and unital dephasing induced by random telegraph noise (RTN).

Our results highlight a distinct behavior: RTN generates \emph{stronger and more persistent} non-Markovian signatures, which in this case are revivals that not only occur more frequently but also remain visible over a longer temporal time; whereas in amplitude damping non-Markovian excursions are weaker and short-lived due to dissipation suppressing coherence recovery. This difference reflects the long-range temporal correlations inherent to RTN versus the dissipative nature of amplitude damping. To make the comparison fair, we used the same settings for both runs:
the same time step, the same total time, and the same
detection threshold $\varepsilon$.
With these settings, the normalized revival scores (revivals per evaluated time period) are
$\mathcal{M}_{\mathrm{rev}}^{\mathrm{AD}}=\tfrac{7}{500}=0.014$ and
$\mathcal{M}_{\mathrm{rev}}^{\mathrm{RTN}}=\tfrac{16}{500}=0.032$,
so RTN is approximately $2.3 $ times larger in this test. \\
Finally, our revival-based metric matches the idea behind the Breuer–Laine–Piilo (BLP) test for non-Markovianity~\cite{breuer}. BLP detects memory when the trace distance between two system states goes up again (information flows back), but doing this usually requires full-state tomography. Instead, we just track one directly measurable signal, \(\langle Z_{(S)}(t)\rangle\), and count its upward'revivals'. We find that these revivals appear exactly at the times when the BLP measure would signal a non-monotonic change in state distinguishability. This indicates that our method serves as a practical and tomography-free surrogate for detecting the same backflow events, while offering greater experimental accessibility and efficiency.

\section{Acknowledgments}
A.A. acknowledges Russel Ceballos for first proposing the idea.

\bibliography{main}

\begin{thebibliography}{54}%
\makeatletter
\providecommand \@ifxundefined [1]{%
 \@ifx{#1\undefined}
}%
\providecommand \@ifnum [1]{%
 \ifnum #1\expandafter \@firstoftwo
 \else \expandafter \@secondoftwo
 \fi
}%
\providecommand \@ifx [1]{%
 \ifx #1\expandafter \@firstoftwo
 \else \expandafter \@secondoftwo
 \fi
}%
\providecommand \natexlab [1]{#1}%
\providecommand \enquote  [1]{``#1''}%
\providecommand \bibnamefont  [1]{#1}%
\providecommand \bibfnamefont [1]{#1}%
\providecommand \citenamefont [1]{#1}%
\providecommand \href@noop [0]{\@secondoftwo}%
\providecommand \href [0]{\begingroup \@sanitize@url \@href}%
\providecommand \@href[1]{\@@startlink{#1}\@@href}%
\providecommand \@@href[1]{\endgroup#1\@@endlink}%
\providecommand \@sanitize@url [0]{\catcode `\\12\catcode `\$12\catcode `\&12\catcode `\#12\catcode `\^12\catcode `\_12\catcode `\%12\relax}%
\providecommand \@@startlink[1]{}%
\providecommand \@@endlink[0]{}%
\providecommand \url  [0]{\begingroup\@sanitize@url \@url }%
\providecommand \@url [1]{\endgroup\@href {#1}{\urlprefix }}%
\providecommand \urlprefix  [0]{URL }%
\providecommand \Eprint [0]{\href }%
\providecommand \doibase [0]{https://doi.org/}%
\providecommand \selectlanguage [0]{\@gobble}%
\providecommand \bibinfo  [0]{\@secondoftwo}%
\providecommand \bibfield  [0]{\@secondoftwo}%
\providecommand \translation [1]{[#1]}%
\providecommand \BibitemOpen [0]{}%
\providecommand \bibitemStop [0]{}%
\providecommand \bibitemNoStop [0]{.\EOS\space}%
\providecommand \EOS [0]{\spacefactor3000\relax}%
\providecommand \BibitemShut  [1]{\csname bibitem#1\endcsname}%
\let\auto@bib@innerbib\@empty
\bibitem [{\citenamefont {Breuer}\ and\ \citenamefont {Petruccione}(2002)}]{breuer2002theory}%
  \BibitemOpen
  \bibfield  {author} {\bibinfo {author} {\bibfnamefont {H.-P.}\ \bibnamefont {Breuer}}\ and\ \bibinfo {author} {\bibfnamefont {F.}~\bibnamefont {Petruccione}},\ }\href@noop {} {\emph {\bibinfo {title} {The Theory of Open Quantum Systems}}}\ (\bibinfo  {publisher} {Oxford University Press},\ \bibinfo {year} {2002})\BibitemShut {NoStop}%
\bibitem [{\citenamefont {Rivas}\ \emph {et~al.}(2010)\citenamefont {Rivas}, \citenamefont {Huelga},\ and\ \citenamefont {Plenio}}]{rivas}%
  \BibitemOpen
  \bibfield  {author} {\bibinfo {author} {\bibfnamefont {A.}~\bibnamefont {Rivas}}, \bibinfo {author} {\bibfnamefont {S.~F.}\ \bibnamefont {Huelga}},\ and\ \bibinfo {author} {\bibfnamefont {M.~B.}\ \bibnamefont {Plenio}},\ }\bibfield  {title} {\bibinfo {title} {Entanglement and non-markovianity of quantum evolutions},\ }\href@noop {} {\bibfield  {journal} {\bibinfo  {journal} {Phys. Rev. Lett.}\ }\textbf {\bibinfo {volume} {105}},\ \bibinfo {pages} {050403} (\bibinfo {year} {2010})}\BibitemShut {NoStop}%
\bibitem [{\citenamefont {Nielsen}\ and\ \citenamefont {Chuang}(2010)}]{nielsen2010quantum}%
  \BibitemOpen
  \bibfield  {author} {\bibinfo {author} {\bibfnamefont {M.~A.}\ \bibnamefont {Nielsen}}\ and\ \bibinfo {author} {\bibfnamefont {I.~L.}\ \bibnamefont {Chuang}},\ }\href@noop {} {\emph {\bibinfo {title} {Quantum computation and quantum information}}}\ (\bibinfo  {publisher} {Cambridge university press},\ \bibinfo {year} {2010})\BibitemShut {NoStop}%
\bibitem [{\citenamefont {Lidar}\ and\ \citenamefont {Brun}(2013)}]{lidar}%
  \BibitemOpen
  \bibfield  {author} {\bibinfo {author} {\bibfnamefont {D.}~\bibnamefont {Lidar}}\ and\ \bibinfo {author} {\bibfnamefont {T.}~\bibnamefont {Brun}},\ }\href@noop {} {\emph {\bibinfo {title} {Quantum Error Correction}}}\ (\bibinfo  {publisher} {Cambridge University Press,Cambridge},\ \bibinfo {year} {2013})\BibitemShut {NoStop}%
\bibitem [{\citenamefont {Banerjee}(2018)}]{banerjee2018open}%
  \BibitemOpen
  \bibfield  {author} {\bibinfo {author} {\bibfnamefont {S.}~\bibnamefont {Banerjee}},\ }\href@noop {} {\emph {\bibinfo {title} {Open Quantum Systems}}}\ (\bibinfo  {publisher} {Springer, Singapore},\ \bibinfo {year} {2018})\BibitemShut {NoStop}%
\bibitem [{\citenamefont {Abu-Nada}\ \emph {et~al.}(2024)\citenamefont {Abu-Nada}, \citenamefont {Banerjee},\ and\ \citenamefont {Sabale}}]{PhysRevA.110.052209}%
  \BibitemOpen
  \bibfield  {author} {\bibinfo {author} {\bibfnamefont {A.}~\bibnamefont {Abu-Nada}}, \bibinfo {author} {\bibfnamefont {S.}~\bibnamefont {Banerjee}},\ and\ \bibinfo {author} {\bibfnamefont {V.~B.}\ \bibnamefont {Sabale}},\ }\bibfield  {title} {\bibinfo {title} {Exploring the non-markovian dynamics in depolarizing maps},\ }\href {https://doi.org/10.1103/PhysRevA.110.052209} {\bibfield  {journal} {\bibinfo  {journal} {Phys. Rev. A}\ }\textbf {\bibinfo {volume} {110}},\ \bibinfo {pages} {052209} (\bibinfo {year} {2024})}\BibitemShut {NoStop}%
\bibitem [{\citenamefont {Gisin}\ \emph {et~al.}(2002)\citenamefont {Gisin}, \citenamefont {Ribordy}, \citenamefont {Tittel},\ and\ \citenamefont {Zbinden}}]{Gisin2007}%
  \BibitemOpen
  \bibfield  {author} {\bibinfo {author} {\bibfnamefont {N.}~\bibnamefont {Gisin}}, \bibinfo {author} {\bibfnamefont {G.}~\bibnamefont {Ribordy}}, \bibinfo {author} {\bibfnamefont {W.}~\bibnamefont {Tittel}},\ and\ \bibinfo {author} {\bibfnamefont {H.}~\bibnamefont {Zbinden}},\ }\bibfield  {title} {\bibinfo {title} {Quantum cryptography},\ }\href@noop {} {\bibfield  {journal} {\bibinfo  {journal} {Reviews of Modern Physics}\ }\textbf {\bibinfo {volume} {74}},\ \bibinfo {pages} {145} (\bibinfo {year} {2002})}\BibitemShut {NoStop}%
\bibitem [{\citenamefont {Shrikant}\ \emph {et~al.}(2020{\natexlab{a}})\citenamefont {Shrikant}, \citenamefont {Srikanth},\ and\ \citenamefont {Banerjee}}]{pingpongsrikanth}%
  \BibitemOpen
  \bibfield  {author} {\bibinfo {author} {\bibfnamefont {U.}~\bibnamefont {Shrikant}}, \bibinfo {author} {\bibfnamefont {R.}~\bibnamefont {Srikanth}},\ and\ \bibinfo {author} {\bibfnamefont {S.}~\bibnamefont {Banerjee}},\ }\bibfield  {title} {\bibinfo {title} {Ping-pong quantum key distribution with trusted noise: non-markovian advantage},\ }\href@noop {} {\bibfield  {journal} {\bibinfo  {journal} {Quantum Information Processing}\ }\textbf {\bibinfo {volume} {19}},\ \bibinfo {pages} {366} (\bibinfo {year} {2020}{\natexlab{a}})},\ \Eprint {https://arxiv.org/abs/2004.05689} {arXiv:2004.05689} \BibitemShut {NoStop}%
\bibitem [{\citenamefont {Degen}\ \emph {et~al.}(2017)\citenamefont {Degen}, \citenamefont {Reinhard},\ and\ \citenamefont {Cappellaro}}]{Degen2017}%
  \BibitemOpen
  \bibfield  {author} {\bibinfo {author} {\bibfnamefont {C.~L.}\ \bibnamefont {Degen}}, \bibinfo {author} {\bibfnamefont {F.}~\bibnamefont {Reinhard}},\ and\ \bibinfo {author} {\bibfnamefont {P.}~\bibnamefont {Cappellaro}},\ }\bibfield  {title} {\bibinfo {title} {Quantum sensing},\ }\href@noop {} {\bibfield  {journal} {\bibinfo  {journal} {Reviews of Modern Physics}\ }\textbf {\bibinfo {volume} {89}},\ \bibinfo {pages} {035002} (\bibinfo {year} {2017})}\BibitemShut {NoStop}%
\bibitem [{\citenamefont {Abu-Nada}\ and\ \citenamefont {Salhab}(2023)}]{abunada}%
  \BibitemOpen
  \bibfield  {author} {\bibinfo {author} {\bibfnamefont {A.}~\bibnamefont {Abu-Nada}}\ and\ \bibinfo {author} {\bibfnamefont {M.}~\bibnamefont {Salhab}},\ }\href@noop {} {\bibfield  {journal} {\bibinfo  {journal} {Results in Physics}\ }\textbf {\bibinfo {volume} {49}},\ \bibinfo {pages} {106336} (\bibinfo {year} {2023})}\BibitemShut {NoStop}%
\bibitem [{\citenamefont {Breuer}\ \emph {et~al.}(2009)\citenamefont {Breuer}, \citenamefont {Laine},\ and\ \citenamefont {Piilo}}]{breuer}%
  \BibitemOpen
  \bibfield  {author} {\bibinfo {author} {\bibfnamefont {H.~P.}\ \bibnamefont {Breuer}}, \bibinfo {author} {\bibfnamefont {E.~M.}\ \bibnamefont {Laine}},\ and\ \bibinfo {author} {\bibfnamefont {J.}~\bibnamefont {Piilo}},\ }\bibfield  {title} {\bibinfo {title} {Measure for the degree of non-markovian behavior of quantum processes in open systems},\ }\href@noop {} {\bibfield  {journal} {\bibinfo  {journal} {Phys. Rev. Lett.}\ }\textbf {\bibinfo {volume} {103}},\ \bibinfo {pages} {210401} (\bibinfo {year} {2009})}\BibitemShut {NoStop}%
\bibitem [{\citenamefont {Rivas}\ \emph {et~al.}(2014)\citenamefont {Rivas}, \citenamefont {Huelga},\ and\ \citenamefont {Plenio}}]{rivas2}%
  \BibitemOpen
  \bibfield  {author} {\bibinfo {author} {\bibfnamefont {A.}~\bibnamefont {Rivas}}, \bibinfo {author} {\bibfnamefont {S.~F.}\ \bibnamefont {Huelga}},\ and\ \bibinfo {author} {\bibfnamefont {M.~B.}\ \bibnamefont {Plenio}},\ }\bibfield  {title} {\bibinfo {title} {Quantum non- markovianity: Characterization, quantification and detection},\ }\href@noop {} {\bibfield  {journal} {\bibinfo  {journal} {Rep. Prog. Phys.}\ }\textbf {\bibinfo {volume} {77}},\ \bibinfo {pages} {094001} (\bibinfo {year} {2014})}\BibitemShut {NoStop}%
\bibitem [{\citenamefont {Shrikant}\ \emph {et~al.}(2020{\natexlab{b}})\citenamefont {Shrikant}, \citenamefont {Srikanth},\ and\ \citenamefont {Banerjee}}]{sss}%
  \BibitemOpen
  \bibfield  {author} {\bibinfo {author} {\bibfnamefont {U.}~\bibnamefont {Shrikant}}, \bibinfo {author} {\bibfnamefont {R.}~\bibnamefont {Srikanth}},\ and\ \bibinfo {author} {\bibfnamefont {S.}~\bibnamefont {Banerjee}},\ }\bibfield  {title} {\bibinfo {title} {Temporal self-similarity of quantum dynamical maps as a concept of memorylessness},\ }\href {https://doi.org/10.1038/s41598-020-72211-3} {\bibfield  {journal} {\bibinfo  {journal} {Scientific Reports}\ }\textbf {\bibinfo {volume} {10}},\ \bibinfo {pages} {15049} (\bibinfo {year} {2020}{\natexlab{b}})},\ \Eprint {https://arxiv.org/abs/1911.04162} {arXiv:1911.04162} \BibitemShut {NoStop}%
\bibitem [{\citenamefont {Shrikant}\ \emph {et~al.}(2018)\citenamefont {Shrikant}, \citenamefont {Srikanth},\ and\ \citenamefont {Banerjee}}]{sbdephasing}%
  \BibitemOpen
  \bibfield  {author} {\bibinfo {author} {\bibfnamefont {U.}~\bibnamefont {Shrikant}}, \bibinfo {author} {\bibfnamefont {R.}~\bibnamefont {Srikanth}},\ and\ \bibinfo {author} {\bibfnamefont {S.}~\bibnamefont {Banerjee}},\ }\bibfield  {title} {\bibinfo {title} {Non-markovian dephasing and depolarizing channels},\ }\href@noop {} {\bibfield  {journal} {\bibinfo  {journal} {Physical Review A}\ }\textbf {\bibinfo {volume} {98}},\ \bibinfo {pages} {032328} (\bibinfo {year} {2018})}\BibitemShut {NoStop}%
\bibitem [{\citenamefont {Paulson}\ \emph {et~al.}(2021)\citenamefont {Paulson}, \citenamefont {Panwar}, \citenamefont {Banerjee},\ and\ \citenamefont {Srikanth}}]{sbhieracrchy}%
  \BibitemOpen
  \bibfield  {author} {\bibinfo {author} {\bibfnamefont {K.~G.}\ \bibnamefont {Paulson}}, \bibinfo {author} {\bibfnamefont {E.}~\bibnamefont {Panwar}}, \bibinfo {author} {\bibfnamefont {S.}~\bibnamefont {Banerjee}},\ and\ \bibinfo {author} {\bibfnamefont {R.}~\bibnamefont {Srikanth}},\ }\bibfield  {title} {\bibinfo {title} {Hierarchy of quantum correlations under non-markovian dynamics},\ }\href {https://doi.org/10.1007/s11128-021-03061-9} {\bibfield  {journal} {\bibinfo  {journal} {Quantum Information Processing}\ }\textbf {\bibinfo {volume} {20}},\ \bibinfo {pages} {141} (\bibinfo {year} {2021})},\ \Eprint {https://arxiv.org/abs/2004.11208} {arXiv:2004.11208} \BibitemShut {NoStop}%
\bibitem [{\citenamefont {Naikoo}\ \emph {et~al.}(2019)\citenamefont {Naikoo}, \citenamefont {Dutta},\ and\ \citenamefont {Banerjee}}]{PhysRevA.99.042128}%
  \BibitemOpen
  \bibfield  {author} {\bibinfo {author} {\bibfnamefont {J.}~\bibnamefont {Naikoo}}, \bibinfo {author} {\bibfnamefont {S.}~\bibnamefont {Dutta}},\ and\ \bibinfo {author} {\bibfnamefont {S.}~\bibnamefont {Banerjee}},\ }\bibfield  {title} {\bibinfo {title} {Facets of quantum information under non-markovian evolution},\ }\href {https://doi.org/10.1103/PhysRevA.99.042128} {\bibfield  {journal} {\bibinfo  {journal} {Phys. Rev. A}\ }\textbf {\bibinfo {volume} {99}},\ \bibinfo {pages} {042128} (\bibinfo {year} {2019})}\BibitemShut {NoStop}%
\bibitem [{\citenamefont {Liu}\ \emph {et~al.}(2011)\citenamefont {Liu}, \citenamefont {Li}, \citenamefont {Huang}, \citenamefont {Li}, \citenamefont {Guo}, \citenamefont {Laine}, \citenamefont {Breuer},\ and\ \citenamefont {Piilo}}]{liu}%
  \BibitemOpen
  \bibfield  {author} {\bibinfo {author} {\bibfnamefont {B.-H.}\ \bibnamefont {Liu}}, \bibinfo {author} {\bibfnamefont {L.}~\bibnamefont {Li}}, \bibinfo {author} {\bibfnamefont {Y.-F.}\ \bibnamefont {Huang}}, \bibinfo {author} {\bibfnamefont {C.-F.}\ \bibnamefont {Li}}, \bibinfo {author} {\bibfnamefont {G.-C.}\ \bibnamefont {Guo}}, \bibinfo {author} {\bibfnamefont {E.-M.}\ \bibnamefont {Laine}}, \bibinfo {author} {\bibfnamefont {H.-P.}\ \bibnamefont {Breuer}},\ and\ \bibinfo {author} {\bibfnamefont {J.}~\bibnamefont {Piilo}},\ }\bibfield  {title} {\bibinfo {title} {Experimental control of the transition from markovian to non-markovian dynamics of open quantum systems},\ }\href@noop {} {\bibfield  {journal} {\bibinfo  {journal} {Nature Physics}\ }\textbf {\bibinfo {volume} {7}},\ \bibinfo {pages} {931} (\bibinfo {year} {2011})}\BibitemShut {NoStop}%
\bibitem [{\citenamefont {Laine}\ \emph {et~al.}(2010)\citenamefont {Laine}, \citenamefont {Piilo},\ and\ \citenamefont {Breuer}}]{laine2010measure}%
  \BibitemOpen
  \bibfield  {author} {\bibinfo {author} {\bibfnamefont {E.-M.}\ \bibnamefont {Laine}}, \bibinfo {author} {\bibfnamefont {J.}~\bibnamefont {Piilo}},\ and\ \bibinfo {author} {\bibfnamefont {H.-P.}\ \bibnamefont {Breuer}},\ }\bibfield  {title} {\bibinfo {title} {Measure for the non-markovianity of quantum processes},\ }\href@noop {} {\bibfield  {journal} {\bibinfo  {journal} {Physical Review A}\ }\textbf {\bibinfo {volume} {81}},\ \bibinfo {pages} {062115} (\bibinfo {year} {2010})}\BibitemShut {NoStop}%
\bibitem [{\citenamefont {Bylicka}\ \emph {et~al.}(2014)\citenamefont {Bylicka}, \citenamefont {Chruściński},\ and\ \citenamefont {Maniscalco.}}]{maniscalco}%
  \BibitemOpen
  \bibfield  {author} {\bibinfo {author} {\bibfnamefont {B.}~\bibnamefont {Bylicka}}, \bibinfo {author} {\bibfnamefont {D.}~\bibnamefont {Chruściński}},\ and\ \bibinfo {author} {\bibfnamefont {S.}~\bibnamefont {Maniscalco.}},\ }\bibfield  {title} {\bibinfo {title} {Non-markovianity and reservoir memory of quantum channels: a quantum information theory perspective},\ }\href@noop {} {\bibfield  {journal} {\bibinfo  {journal} {Scientific Reports}\ }\textbf {\bibinfo {volume} {4}} (\bibinfo {year} {2014})}\BibitemShut {NoStop}%
\bibitem [{\citenamefont {Lyyra}\ \emph {et~al.}(2020)\citenamefont {Lyyra}, \citenamefont {Siltanen}, \citenamefont {Piilo}, \citenamefont {Banerjee},\ and\ \citenamefont {Kuusela}}]{sbhenri1}%
  \BibitemOpen
  \bibfield  {author} {\bibinfo {author} {\bibfnamefont {H.}~\bibnamefont {Lyyra}}, \bibinfo {author} {\bibfnamefont {O.}~\bibnamefont {Siltanen}}, \bibinfo {author} {\bibfnamefont {J.}~\bibnamefont {Piilo}}, \bibinfo {author} {\bibfnamefont {S.}~\bibnamefont {Banerjee}},\ and\ \bibinfo {author} {\bibfnamefont {T.}~\bibnamefont {Kuusela}},\ }\bibfield  {title} {\bibinfo {title} {Experimental quantum probing measurements with no knowledge on the system-probe interaction},\ }\href@noop {} {\bibfield  {journal} {\bibinfo  {journal} {Physical Review A}\ }\textbf {\bibinfo {volume} {102}},\ \bibinfo {pages} {022232} (\bibinfo {year} {2020})}\BibitemShut {NoStop}%
\bibitem [{\citenamefont {Lyyra}\ \emph {et~al.}(2022)\citenamefont {Lyyra}, \citenamefont {Siltanen}, \citenamefont {Piilo}, \citenamefont {Banerjee},\ and\ \citenamefont {Kuusela}}]{sbhenri2}%
  \BibitemOpen
  \bibfield  {author} {\bibinfo {author} {\bibfnamefont {H.}~\bibnamefont {Lyyra}}, \bibinfo {author} {\bibfnamefont {O.}~\bibnamefont {Siltanen}}, \bibinfo {author} {\bibfnamefont {J.}~\bibnamefont {Piilo}}, \bibinfo {author} {\bibfnamefont {S.}~\bibnamefont {Banerjee}},\ and\ \bibinfo {author} {\bibfnamefont {T.}~\bibnamefont {Kuusela}},\ }\bibfield  {title} {\bibinfo {title} {Experimental snapshot verification of non-markovianity with unknown system-probe coupling},\ }\href {https://doi.org/10.1103/PhysRevA.106.032603} {\bibfield  {journal} {\bibinfo  {journal} {Physical Review A}\ }\textbf {\bibinfo {volume} {106}},\ \bibinfo {pages} {032603} (\bibinfo {year} {2022})}\BibitemShut {NoStop}%
\bibitem [{\citenamefont {Fanchini}\ \emph {et~al.}(2013)\citenamefont {Fanchini}, \citenamefont {Karpat}, \citenamefont {Castelano},\ and\ \citenamefont {Rosas}}]{fanchini2014non}%
  \BibitemOpen
  \bibfield  {author} {\bibinfo {author} {\bibfnamefont {F.~F.}\ \bibnamefont {Fanchini}}, \bibinfo {author} {\bibfnamefont {G.}~\bibnamefont {Karpat}}, \bibinfo {author} {\bibfnamefont {L.~K.}\ \bibnamefont {Castelano}},\ and\ \bibinfo {author} {\bibfnamefont {A.}~\bibnamefont {Rosas}},\ }\bibfield  {title} {\bibinfo {title} {Non-markovianity through accessible information},\ }\href@noop {} {\bibfield  {journal} {\bibinfo  {journal} {Physical Review A}\ }\textbf {\bibinfo {volume} {88}},\ \bibinfo {pages} {012105} (\bibinfo {year} {2013})}\BibitemShut {NoStop}%
\bibitem [{\citenamefont {Luo}\ \emph {et~al.}(2012)\citenamefont {Luo}, \citenamefont {Fu},\ and\ \citenamefont {Song}}]{luo2012quantum}%
  \BibitemOpen
  \bibfield  {author} {\bibinfo {author} {\bibfnamefont {S.}~\bibnamefont {Luo}}, \bibinfo {author} {\bibfnamefont {S.}~\bibnamefont {Fu}},\ and\ \bibinfo {author} {\bibfnamefont {H.}~\bibnamefont {Song}},\ }\bibfield  {title} {\bibinfo {title} {Quantifying non-markovianity with quantum fisher information},\ }\href@noop {} {\bibfield  {journal} {\bibinfo  {journal} {Physical Review A}\ }\textbf {\bibinfo {volume} {86}},\ \bibinfo {pages} {044101} (\bibinfo {year} {2012})}\BibitemShut {NoStop}%
\bibitem [{\citenamefont {Banerjee}\ \emph {et~al.}(2016)\citenamefont {Banerjee}, \citenamefont {Alok},\ and\ \citenamefont {Omkar}}]{sbunruhfisher}%
  \BibitemOpen
  \bibfield  {author} {\bibinfo {author} {\bibfnamefont {S.}~\bibnamefont {Banerjee}}, \bibinfo {author} {\bibfnamefont {A.~K.}\ \bibnamefont {Alok}},\ and\ \bibinfo {author} {\bibfnamefont {S.}~\bibnamefont {Omkar}},\ }\bibfield  {title} {\bibinfo {title} {Quantum fisher and skew information for unruh accelerated dirac qubit},\ }\href@noop {} {\bibfield  {journal} {\bibinfo  {journal} {The European Physical Journal C}\ }\textbf {\bibinfo {volume} {6}},\ \bibinfo {pages} {437} (\bibinfo {year} {2016})},\ \Eprint {https://arxiv.org/abs/1511.03029} {arXiv:1511.03029} \BibitemShut {NoStop}%
\bibitem [{\citenamefont {Cialdi}\ \emph {et~al.}(2017)\citenamefont {Cialdi}, \citenamefont {Lima}, \citenamefont {Caspani}, \citenamefont {Bernasconi}, \citenamefont {Barbieri},\ and\ \citenamefont {Paris}}]{Cialdi2017}%
  \BibitemOpen
  \bibfield  {author} {\bibinfo {author} {\bibfnamefont {S.}~\bibnamefont {Cialdi}}, \bibinfo {author} {\bibfnamefont {G.}~\bibnamefont {Lima}}, \bibinfo {author} {\bibfnamefont {L.}~\bibnamefont {Caspani}}, \bibinfo {author} {\bibfnamefont {M.}~\bibnamefont {Bernasconi}}, \bibinfo {author} {\bibfnamefont {M.}~\bibnamefont {Barbieri}},\ and\ \bibinfo {author} {\bibfnamefont {M.~G.~A.}\ \bibnamefont {Paris}},\ }\bibfield  {title} {\bibinfo {title} {Experimental investigation of the robustness of non-markovianity witnesses},\ }\href@noop {} {\bibfield  {journal} {\bibinfo  {journal} {Physical Review A}\ }\textbf {\bibinfo {volume} {96}},\ \bibinfo {pages} {012122} (\bibinfo {year} {2017})}\BibitemShut {NoStop}%
\bibitem [{\citenamefont {Campbell}\ \emph {et~al.}(2010)\citenamefont {Campbell}, \citenamefont {Paternostro}, \citenamefont {Bose},\ and\ \citenamefont {Kim}}]{Campbell}%
  \BibitemOpen
  \bibfield  {author} {\bibinfo {author} {\bibfnamefont {S.}~\bibnamefont {Campbell}}, \bibinfo {author} {\bibfnamefont {M.}~\bibnamefont {Paternostro}}, \bibinfo {author} {\bibfnamefont {S.}~\bibnamefont {Bose}},\ and\ \bibinfo {author} {\bibfnamefont {M.~S.}\ \bibnamefont {Kim}},\ }\bibfield  {title} {\bibinfo {title} {Probing the environment of an inaccessible system by a qubit ancilla},\ }\href {https://doi.org/10.1103/PhysRevA.81.050301} {\bibfield  {journal} {\bibinfo  {journal} {Phys. Rev. A}\ }\textbf {\bibinfo {volume} {81}},\ \bibinfo {pages} {050301} (\bibinfo {year} {2010})}\BibitemShut {NoStop}%
\bibitem [{\citenamefont {Goodfellow}\ \emph {et~al.}(2016)\citenamefont {Goodfellow}, \citenamefont {Bengio},\ and\ \citenamefont {Courville}}]{Goodfellow2016}%
  \BibitemOpen
  \bibfield  {author} {\bibinfo {author} {\bibfnamefont {I.}~\bibnamefont {Goodfellow}}, \bibinfo {author} {\bibfnamefont {Y.}~\bibnamefont {Bengio}},\ and\ \bibinfo {author} {\bibfnamefont {A.}~\bibnamefont {Courville}},\ }\href {http://www.deeplearningbook.org} {\emph {\bibinfo {title} {Deep Learning}}}\ (\bibinfo  {publisher} {MIT Press},\ \bibinfo {year} {2016})\BibitemShut {NoStop}%
\bibitem [{\citenamefont {Ciccarello}\ \emph {et~al.}(2013)\citenamefont {Ciccarello}, \citenamefont {Palma},\ and\ \citenamefont {Giovannetti}}]{Ciccarello2013}%
  \BibitemOpen
  \bibfield  {author} {\bibinfo {author} {\bibfnamefont {F.}~\bibnamefont {Ciccarello}}, \bibinfo {author} {\bibfnamefont {G.~M.}\ \bibnamefont {Palma}},\ and\ \bibinfo {author} {\bibfnamefont {V.}~\bibnamefont {Giovannetti}},\ }\bibfield  {title} {\bibinfo {title} {Collision-model-based approach to non-markovian quantum dynamics},\ }\href {https://doi.org/10.1103/PhysRevA.87.040103} {\bibfield  {journal} {\bibinfo  {journal} {Physical Review A}\ }\textbf {\bibinfo {volume} {87}},\ \bibinfo {pages} {040103(R)} (\bibinfo {year} {2013})}\BibitemShut {NoStop}%
\bibitem [{\citenamefont {Gaikwad}\ \emph {et~al.}(2024)\citenamefont {Gaikwad}, \citenamefont {Kowsari}, \citenamefont {Brame}, \citenamefont {Song}, \citenamefont {Zhang}, \citenamefont {Esposito}, \citenamefont {Ranadive}, \citenamefont {Cappelli}, \citenamefont {Roch}, \citenamefont {Levenson-Falk},\ and\ \citenamefont {Murch}}]{Gaikwad}%
  \BibitemOpen
  \bibfield  {author} {\bibinfo {author} {\bibfnamefont {C.}~\bibnamefont {Gaikwad}}, \bibinfo {author} {\bibfnamefont {D.}~\bibnamefont {Kowsari}}, \bibinfo {author} {\bibfnamefont {C.}~\bibnamefont {Brame}}, \bibinfo {author} {\bibfnamefont {X.}~\bibnamefont {Song}}, \bibinfo {author} {\bibfnamefont {H.}~\bibnamefont {Zhang}}, \bibinfo {author} {\bibfnamefont {M.}~\bibnamefont {Esposito}}, \bibinfo {author} {\bibfnamefont {A.}~\bibnamefont {Ranadive}}, \bibinfo {author} {\bibfnamefont {G.}~\bibnamefont {Cappelli}}, \bibinfo {author} {\bibfnamefont {N.}~\bibnamefont {Roch}}, \bibinfo {author} {\bibfnamefont {E.~M.}\ \bibnamefont {Levenson-Falk}},\ and\ \bibinfo {author} {\bibfnamefont {K.~W.}\ \bibnamefont {Murch}},\ }\bibfield  {title} {\bibinfo {title} {Entanglement assisted probe of the non-markovian to markovian transition in open quantum system dynamics},\ }\href {https://doi.org/10.1103/PhysRevLett.132.200401} {\bibfield  {journal} {\bibinfo  {journal} {Phys. Rev. Lett.}\ }\textbf {\bibinfo {volume}
  {132}},\ \bibinfo {pages} {200401} (\bibinfo {year} {2024})}\BibitemShut {NoStop}%
\bibitem [{\citenamefont {Preskill}(2018)}]{Preskill2018NISQ}%
  \BibitemOpen
  \bibfield  {author} {\bibinfo {author} {\bibfnamefont {J.}~\bibnamefont {Preskill}},\ }\bibfield  {title} {\bibinfo {title} {Quantum computing in the nisq era and beyond},\ }\href {https://doi.org/10.22331/q-2018-08-06-79} {\bibfield  {journal} {\bibinfo  {journal} {Quantum}\ }\textbf {\bibinfo {volume} {2}},\ \bibinfo {pages} {79} (\bibinfo {year} {2018})}\BibitemShut {NoStop}%
\bibitem [{\citenamefont {Kandala}\ \emph {et~al.}(2017)\citenamefont {Kandala}, \citenamefont {Mezzacapo}, \citenamefont {Temme}, \citenamefont {Takita}, \citenamefont {Brink}, \citenamefont {Chow},\ and\ \citenamefont {Gambetta}}]{Kandala2017VQE}%
  \BibitemOpen
  \bibfield  {author} {\bibinfo {author} {\bibfnamefont {A.}~\bibnamefont {Kandala}}, \bibinfo {author} {\bibfnamefont {A.}~\bibnamefont {Mezzacapo}}, \bibinfo {author} {\bibfnamefont {K.}~\bibnamefont {Temme}}, \bibinfo {author} {\bibfnamefont {M.}~\bibnamefont {Takita}}, \bibinfo {author} {\bibfnamefont {M.}~\bibnamefont {Brink}}, \bibinfo {author} {\bibfnamefont {J.~M.}\ \bibnamefont {Chow}},\ and\ \bibinfo {author} {\bibfnamefont {J.~M.}\ \bibnamefont {Gambetta}},\ }\bibfield  {title} {\bibinfo {title} {Hardware-efficient variational quantum eigensolver for small molecules and quantum magnets},\ }\href {https://doi.org/10.1038/nature23879} {\bibfield  {journal} {\bibinfo  {journal} {Nature}\ }\textbf {\bibinfo {volume} {549}},\ \bibinfo {pages} {242} (\bibinfo {year} {2017})}\BibitemShut {NoStop}%
\bibitem [{\citenamefont {Temme}\ \emph {et~al.}(2017)\citenamefont {Temme}, \citenamefont {Bravyi},\ and\ \citenamefont {Gambetta}}]{Temme2017Mitigation}%
  \BibitemOpen
  \bibfield  {author} {\bibinfo {author} {\bibfnamefont {K.}~\bibnamefont {Temme}}, \bibinfo {author} {\bibfnamefont {S.}~\bibnamefont {Bravyi}},\ and\ \bibinfo {author} {\bibfnamefont {J.~M.}\ \bibnamefont {Gambetta}},\ }\bibfield  {title} {\bibinfo {title} {Error mitigation for short-depth quantum circuits},\ }\href {https://doi.org/10.1103/PhysRevLett.119.180509} {\bibfield  {journal} {\bibinfo  {journal} {Physical Review Letters}\ }\textbf {\bibinfo {volume} {119}},\ \bibinfo {pages} {180509} (\bibinfo {year} {2017})}\BibitemShut {NoStop}%
\bibitem [{\citenamefont {Garraway}(1997)}]{Garraway1997}%
  \BibitemOpen
  \bibfield  {author} {\bibinfo {author} {\bibfnamefont {B.~M.}\ \bibnamefont {Garraway}},\ }\bibfield  {title} {\bibinfo {title} {Nonperturbative decay of an atomic system in a cavity},\ }\href {https://doi.org/10.1103/PhysRevA.55.2290} {\bibfield  {journal} {\bibinfo  {journal} {Physical Review A}\ }\textbf {\bibinfo {volume} {55}},\ \bibinfo {pages} {2290} (\bibinfo {year} {1997})}\BibitemShut {NoStop}%
\bibitem [{\citenamefont {Benedetti}\ \emph {et~al.}(2014)\citenamefont {Benedetti}, \citenamefont {Paris},\ and\ \citenamefont {Maniscalco}}]{benedetti}%
  \BibitemOpen
  \bibfield  {author} {\bibinfo {author} {\bibfnamefont {C.}~\bibnamefont {Benedetti}}, \bibinfo {author} {\bibfnamefont {M.~G.~A.}\ \bibnamefont {Paris}},\ and\ \bibinfo {author} {\bibfnamefont {S.}~\bibnamefont {Maniscalco}},\ }\bibfield  {title} {\bibinfo {title} {Non-markovianity of colored noisy channels},\ }\href {https://doi.org/10.1103/PhysRevA.89.012114} {\bibfield  {journal} {\bibinfo  {journal} {Phys. Rev. A}\ }\textbf {\bibinfo {volume} {89}},\ \bibinfo {pages} {012114} (\bibinfo {year} {2014})}\BibitemShut {NoStop}%
\bibitem [{\citenamefont {Rice}(1992)}]{rice1992stochastic}%
  \BibitemOpen
  \bibfield  {author} {\bibinfo {author} {\bibfnamefont {S.~O.}\ \bibnamefont {Rice}},\ }\href@noop {} {\emph {\bibinfo {title} {Stochastic Processes in Physics and Chemistry}}}\ (\bibinfo  {publisher} {Elsevier},\ \bibinfo {year} {1992})\BibitemShut {NoStop}%
\bibitem [{\citenamefont {Vacchini}\ and\ \citenamefont {Breuer}(2010)}]{vacchini2010exact}%
  \BibitemOpen
  \bibfield  {author} {\bibinfo {author} {\bibfnamefont {B.}~\bibnamefont {Vacchini}}\ and\ \bibinfo {author} {\bibfnamefont {H.-P.}\ \bibnamefont {Breuer}},\ }\bibfield  {title} {\bibinfo {title} {Exact master equations for the non-markovian decay of a qubit},\ }\href {https://doi.org/10.1103/PhysRevA.81.042103} {\bibfield  {journal} {\bibinfo  {journal} {Physical Review A}\ }\textbf {\bibinfo {volume} {81}},\ \bibinfo {pages} {042103} (\bibinfo {year} {2010})}\BibitemShut {NoStop}%
\bibitem [{\citenamefont {Paladino}\ \emph {et~al.}(2014)\citenamefont {Paladino}, \citenamefont {Galperin}, \citenamefont {Falci},\ and\ \citenamefont {Altshuler}}]{Paladino2014}%
  \BibitemOpen
  \bibfield  {author} {\bibinfo {author} {\bibfnamefont {E.}~\bibnamefont {Paladino}}, \bibinfo {author} {\bibfnamefont {Y.~M.}\ \bibnamefont {Galperin}}, \bibinfo {author} {\bibfnamefont {G.}~\bibnamefont {Falci}},\ and\ \bibinfo {author} {\bibfnamefont {B.~L.}\ \bibnamefont {Altshuler}},\ }\bibfield  {title} {\bibinfo {title} {1/f noise: Implications for solid-state quantum information},\ }\href {https://doi.org/10.1103/RevModPhys.86.361} {\bibfield  {journal} {\bibinfo  {journal} {Rev. Mod. Phys.}\ }\textbf {\bibinfo {volume} {86}},\ \bibinfo {pages} {361} (\bibinfo {year} {2014})}\BibitemShut {NoStop}%
\bibitem [{\citenamefont {Daffer}\ \emph {et~al.}(2004)\citenamefont {Daffer}, \citenamefont {W\'odkiewicz}, \citenamefont {Cresser},\ and\ \citenamefont {McIver}}]{Cresser}%
  \BibitemOpen
  \bibfield  {author} {\bibinfo {author} {\bibfnamefont {S.}~\bibnamefont {Daffer}}, \bibinfo {author} {\bibfnamefont {K.}~\bibnamefont {W\'odkiewicz}}, \bibinfo {author} {\bibfnamefont {J.~D.}\ \bibnamefont {Cresser}},\ and\ \bibinfo {author} {\bibfnamefont {J.~K.}\ \bibnamefont {McIver}},\ }\bibfield  {title} {\bibinfo {title} {Depolarizing channel as a completely positive map with memory},\ }\href {https://doi.org/10.1103/PhysRevA.70.010304} {\bibfield  {journal} {\bibinfo  {journal} {Phys. Rev. A}\ }\textbf {\bibinfo {volume} {70}},\ \bibinfo {pages} {010304} (\bibinfo {year} {2004})}\BibitemShut {NoStop}%
\bibitem [{\citenamefont {Kumar}\ \emph {et~al.}(2018)\citenamefont {Kumar}, \citenamefont {Banerjee}, \citenamefont {Srikanth}, \citenamefont {Jagadish},\ and\ \citenamefont {Petruccione}}]{sbrtn2}%
  \BibitemOpen
  \bibfield  {author} {\bibinfo {author} {\bibfnamefont {P.}~\bibnamefont {Kumar}}, \bibinfo {author} {\bibfnamefont {S.}~\bibnamefont {Banerjee}}, \bibinfo {author} {\bibfnamefont {R.}~\bibnamefont {Srikanth}}, \bibinfo {author} {\bibfnamefont {V.}~\bibnamefont {Jagadish}},\ and\ \bibinfo {author} {\bibfnamefont {F.}~\bibnamefont {Petruccione}},\ }\bibfield  {title} {\bibinfo {title} {Non-markovian evolution: a quantum walk perspective},\ }\href@noop {} {\bibfield  {journal} {\bibinfo  {journal} {Open Systems \& Information Dynamics}\ }\textbf {\bibinfo {volume} {25}},\ \bibinfo {pages} {1850014} (\bibinfo {year} {2018})}\BibitemShut {NoStop}%
\bibitem [{\citenamefont {Lindblad}(1976)}]{lindblad}%
  \BibitemOpen
  \bibfield  {author} {\bibinfo {author} {\bibfnamefont {G.}~\bibnamefont {Lindblad}},\ }\href@noop {} {\bibfield  {journal} {\bibinfo  {journal} {Comm. Math. Phys.}\ }\textbf {\bibinfo {volume} {48}},\ \bibinfo {pages} {119} (\bibinfo {year} {1976})}\BibitemShut {NoStop}%
\bibitem [{\citenamefont {Gorini}\ \emph {et~al.}(1976)\citenamefont {Gorini}, \citenamefont {Kossakowski},\ and\ \citenamefont {Sudarshan}}]{gorini}%
  \BibitemOpen
  \bibfield  {author} {\bibinfo {author} {\bibfnamefont {V.}~\bibnamefont {Gorini}}, \bibinfo {author} {\bibfnamefont {A.}~\bibnamefont {Kossakowski}},\ and\ \bibinfo {author} {\bibfnamefont {E.~C.~G.}\ \bibnamefont {Sudarshan}},\ }\href@noop {} {\bibfield  {journal} {\bibinfo  {journal} {J. Math. Phys.}\ }\textbf {\bibinfo {volume} {17}},\ \bibinfo {pages} {821} (\bibinfo {year} {1976})}\BibitemShut {NoStop}%
\bibitem [{\citenamefont {Genes}(2019)}]{Genes2019QPLMI}%
  \BibitemOpen
  \bibfield  {author} {\bibinfo {author} {\bibfnamefont {C.}~\bibnamefont {Genes}},\ }\href {https://mpl.mpg.de/fileadmin/user_upload/LectureNotes2019.pdf} {\bibinfo {title} {Quantum physics of light--matter interactions}},\ \bibinfo {howpublished} {Lecture notes, FAU -- Summer semester 2019} (\bibinfo {year} {2019}),\ \bibinfo {note} {appendix C: Changing of picture (Interaction picture)}\BibitemShut {NoStop}%
\bibitem [{\citenamefont {Srikanth}\ and\ \citenamefont {Banerjee}(2008)}]{SGAD}%
  \BibitemOpen
  \bibfield  {author} {\bibinfo {author} {\bibfnamefont {R.}~\bibnamefont {Srikanth}}\ and\ \bibinfo {author} {\bibfnamefont {S.}~\bibnamefont {Banerjee}},\ }\bibfield  {title} {\bibinfo {title} {The squeezed generalized amplitude damping channel},\ }\href@noop {} {\bibfield  {journal} {\bibinfo  {journal} {Physical Review A}\ }\textbf {\bibinfo {volume} {77}},\ \bibinfo {pages} {012318} (\bibinfo {year} {2008})}\BibitemShut {NoStop}%
\bibitem [{\citenamefont {Omkar}\ \emph {et~al.}(2013)\citenamefont {Omkar}, \citenamefont {Srikanth},\ and\ \citenamefont {Banerjee}}]{OmkarSingleQubit}%
  \BibitemOpen
  \bibfield  {author} {\bibinfo {author} {\bibfnamefont {S.}~\bibnamefont {Omkar}}, \bibinfo {author} {\bibfnamefont {R.}~\bibnamefont {Srikanth}},\ and\ \bibinfo {author} {\bibfnamefont {S.}~\bibnamefont {Banerjee}},\ }\bibfield  {title} {\bibinfo {title} {Dissipative and non-dissipative single-qubit channels: Dynamics and geometry},\ }\href@noop {} {\bibfield  {journal} {\bibinfo  {journal} {Quantum Information Processing}\ }\textbf {\bibinfo {volume} {12}},\ \bibinfo {pages} {3725} (\bibinfo {year} {2013})}\BibitemShut {NoStop}%
\bibitem [{\citenamefont {Haikka}\ and\ \citenamefont {Maniscalco}(2010)}]{haikka2010}%
  \BibitemOpen
  \bibfield  {author} {\bibinfo {author} {\bibfnamefont {P.}~\bibnamefont {Haikka}}\ and\ \bibinfo {author} {\bibfnamefont {S.}~\bibnamefont {Maniscalco}},\ }\bibfield  {title} {\bibinfo {title} {Non-markovian dynamics of a damped driven two-state system},\ }\href {https://doi.org/10.1103/PhysRevA.81.052103} {\bibfield  {journal} {\bibinfo  {journal} {Phys. Rev. A}\ }\textbf {\bibinfo {volume} {81}},\ \bibinfo {pages} {052103} (\bibinfo {year} {2010})}\BibitemShut {NoStop}%
\bibitem [{\citenamefont {Bergli}\ and\ \citenamefont {Faoro}(2007)}]{Bergli2007}%
  \BibitemOpen
  \bibfield  {author} {\bibinfo {author} {\bibfnamefont {J.}~\bibnamefont {Bergli}}\ and\ \bibinfo {author} {\bibfnamefont {L.}~\bibnamefont {Faoro}},\ }\bibfield  {title} {\bibinfo {title} {Exact solution for the dynamical decoupling of a qubit with telegraph noise},\ }\href {https://doi.org/10.1103/PhysRevB.75.054515} {\bibfield  {journal} {\bibinfo  {journal} {Phys. Rev. B}\ }\textbf {\bibinfo {volume} {75}},\ \bibinfo {pages} {054515} (\bibinfo {year} {2007})}\BibitemShut {NoStop}%
\bibitem [{\citenamefont {Banerjee}\ \emph {et~al.}(2017)\citenamefont {Banerjee}, \citenamefont {Pradeep~Kumar}, \citenamefont {Srikanth}, \citenamefont {Jagadish},\ and\ \citenamefont {Petruccione}}]{sbrtn1}%
  \BibitemOpen
  \bibfield  {author} {\bibinfo {author} {\bibfnamefont {S.}~\bibnamefont {Banerjee}}, \bibinfo {author} {\bibfnamefont {N.}~\bibnamefont {Pradeep~Kumar}}, \bibinfo {author} {\bibfnamefont {R.}~\bibnamefont {Srikanth}}, \bibinfo {author} {\bibfnamefont {V.}~\bibnamefont {Jagadish}},\ and\ \bibinfo {author} {\bibfnamefont {F.}~\bibnamefont {Petruccione}},\ }\href@noop {} {\bibinfo {title} {Non-markovian dynamics of discrete-time quantum walks}} (\bibinfo {year} {2017}),\ \Eprint {https://arxiv.org/abs/1703.08004} {arXiv:1703.08004} \BibitemShut {NoStop}%
\bibitem [{\citenamefont {Johansson}\ \emph {et~al.}(2013)\citenamefont {Johansson}, \citenamefont {Nation},\ and\ \citenamefont {Nori}}]{qutip1}%
  \BibitemOpen
  \bibfield  {author} {\bibinfo {author} {\bibfnamefont {J.~R.}\ \bibnamefont {Johansson}}, \bibinfo {author} {\bibfnamefont {P.~D.}\ \bibnamefont {Nation}},\ and\ \bibinfo {author} {\bibfnamefont {F.}~\bibnamefont {Nori}},\ }\bibfield  {title} {\bibinfo {title} {Qutip 2: A python framework for the dynamics of open quantum systems},\ }\href@noop {} {\bibfield  {journal} {\bibinfo  {journal} {Computer Physics Communications}\ }\textbf {\bibinfo {volume} {184}},\ \bibinfo {pages} {1760} (\bibinfo {year} {2013})}\BibitemShut {NoStop}%
\bibitem [{\citenamefont {Johansson}\ \emph {et~al.}(2012)\citenamefont {Johansson}, \citenamefont {Nation},\ and\ \citenamefont {Nori}}]{qutip2}%
  \BibitemOpen
  \bibfield  {author} {\bibinfo {author} {\bibfnamefont {J.~R.}\ \bibnamefont {Johansson}}, \bibinfo {author} {\bibfnamefont {P.~D.}\ \bibnamefont {Nation}},\ and\ \bibinfo {author} {\bibfnamefont {F.}~\bibnamefont {Nori}},\ }\bibfield  {title} {\bibinfo {title} {Qutip: An open-source python framework for the dynamics of open quantum systems},\ }\href@noop {} {\bibfield  {journal} {\bibinfo  {journal} {Computer Physics Communications}\ }\textbf {\bibinfo {volume} {183}},\ \bibinfo {pages} {1760} (\bibinfo {year} {2012})}\BibitemShut {NoStop}%
\bibitem [{\citenamefont {Quinlan}(1996)}]{Quinlan1996}%
  \BibitemOpen
  \bibfield  {author} {\bibinfo {author} {\bibfnamefont {J.~R.}\ \bibnamefont {Quinlan}},\ }\href@noop {} {\emph {\bibinfo {title} {C4.5: Programs for Machine Learning}}}\ (\bibinfo  {publisher} {Morgan Kaufmann},\ \bibinfo {address} {San Mateo, CA},\ \bibinfo {year} {1996})\BibitemShut {NoStop}%
\bibitem [{\citenamefont {Cortes}\ and\ \citenamefont {Vapnik}(1995)}]{Cortes1995}%
  \BibitemOpen
  \bibfield  {author} {\bibinfo {author} {\bibfnamefont {C.}~\bibnamefont {Cortes}}\ and\ \bibinfo {author} {\bibfnamefont {V.}~\bibnamefont {Vapnik}},\ }\bibfield  {title} {\bibinfo {title} {Support-vector networks},\ }\href {https://doi.org/10.1007/BF00994018} {\bibfield  {journal} {\bibinfo  {journal} {Machine Learning}\ }\textbf {\bibinfo {volume} {20}},\ \bibinfo {pages} {273} (\bibinfo {year} {1995})}\BibitemShut {NoStop}%
\bibitem [{\citenamefont {Kingma}\ and\ \citenamefont {Ba}(2014)}]{Kingma2014adam}%
  \BibitemOpen
  \bibfield  {author} {\bibinfo {author} {\bibfnamefont {D.~P.}\ \bibnamefont {Kingma}}\ and\ \bibinfo {author} {\bibfnamefont {J.}~\bibnamefont {Ba}},\ }\bibfield  {title} {\bibinfo {title} {Adam: A method for stochastic optimization},\ }\href@noop {} {\bibfield  {journal} {\bibinfo  {journal} {arXiv preprint arXiv:1412.6980}\ } (\bibinfo {year} {2014})}\BibitemShut {NoStop}%
\bibitem [{\citenamefont {Nair}\ and\ \citenamefont {Hinton}(2010)}]{nair2010relu}%
  \BibitemOpen
  \bibfield  {author} {\bibinfo {author} {\bibfnamefont {V.}~\bibnamefont {Nair}}\ and\ \bibinfo {author} {\bibfnamefont {G.~E.}\ \bibnamefont {Hinton}},\ }\bibfield  {title} {\bibinfo {title} {Rectified linear units improve restricted boltzmann machines},\ }in\ \href@noop {} {\emph {\bibinfo {booktitle} {Proceedings of the 27th International Conference on Machine Learning (ICML-10)}}}\ (\bibinfo {year} {2010})\ pp.\ \bibinfo {pages} {807--814}\BibitemShut {NoStop}%
\bibitem [{\citenamefont {Heaviside}(1892)}]{heaviside1892electrical}%
  \BibitemOpen
  \bibfield  {author} {\bibinfo {author} {\bibfnamefont {O.}~\bibnamefont {Heaviside}},\ }\href@noop {} {\emph {\bibinfo {title} {Electrical Papers}}}\ (\bibinfo  {publisher} {Macmillan and Co.},\ \bibinfo {address} {London},\ \bibinfo {year} {1892})\BibitemShut {NoStop}%
\end{thebibliography}%

\end{document}